\documentclass[11pt]{article}
\newcommand{\Comment}[1]{{}}
\usepackage[textwidth = 450 pt, textheight = 630 pt]{geometry}
\usepackage{amssymb,euscript,amsmath,amsfonts}
\usepackage{tikz,tikz-cd,url}
\usetikzlibrary{decorations.pathreplacing,calligraphy}
\usepackage{xcolor}
\usepackage{graphicx}
\usepackage{color}

\definecolor{MyDarkBlue}{rgb}{0.15,0.15,0.45}
\usepackage[linktocpage=true]{hyperref}
\hypersetup{final,
colorlinks=true,
citecolor=MyDarkBlue,
linkcolor=MyDarkBlue,
urlcolor=MyDarkBlue,
pdfauthor={E. Andriolo, N. Lambert, T. Orchard and C. Papageorgakis}
pdftitle={Title},
pdfsubject={hep-th}
}

\usepackage[numbers,sort&compress]{natbib}
\usepackage{hypernat}
\usepackage{accents}
\usepackage{bm}

\newcommand{\be}{\begin{equation}}
\newcommand{\ee}{\end{equation}}
\newcommand{\bea}{\begin{eqnarray}}
\newcommand{\eea}{\end{eqnarray}}

\numberwithin{equation}{section}

\begin{document}

\renewcommand{\thefootnote}{\fnsymbol{footnote}}
 
 \rightline{QMUL-PH-21-50}

   \vspace{1.8truecm}

 \centerline{\LARGE \bf {\sc A Path Integral for the Chiral-Form Partition Function}}
 
 \vspace{1cm}

 \centerline{
   {\large {\bf {\sc E.~Andriolo,$^{\,a}$}\footnote{E-mail address: \href{mailto:e.andriolo@qmul.ac.uk}{\tt e.andriolo@qmul.ac.uk}}\,, {\sc N.~Lambert,$^{\,b}$}\footnote{E-mail address: \href{mailto:neil.lambert@kcl.ac.uk}{\tt neil.lambert@kcl.ac.uk}}\, {\sc T.~Orchard${}^{\,b}$}}\footnote{E-mail address: \href{mailto:tristan.orchard@kcl.ac.uk}{\tt tristan.orchard@kcl.ac.uk}}\, {\sc and C.~Papageorgakis$^{\,a}$}\footnote{E-mail address: \href{mailto:c.papageorgakis@qmul.ac.uk}{\tt c.papageorgakis@qmul.ac.uk}}} }

\vspace{1cm}
\centerline{${}^a${\it Centre for Theoretical Physics,  Department of Physics and Astronomy}} 
\centerline{{\it Queen Mary University of London,  London E1 4NS, UK}} 
\vskip 1cm
\centerline{${}^b${\it Department of Mathematics}}
\centerline{{\it King's College London, London WC2R 2LS, UK}} 

\vspace{1.0truecm}

 
\thispagestyle{empty}

\centerline{\sc Abstract}
\vspace{0.4truecm}
\begin{center}
\begin{minipage}[c]{360pt}{Starting from the recent action  proposed by Sen \cite{Sen:2015nph,Sen:2019qit}, we evaluate the partition function of the compact chiral boson on a two-dimensional torus using a path-integral formulation. Crucially,  we use a Wick-rotation procedure obtained from a complex deformation of the physical spacetime metric. This directly reproduces the expected result including general characteristics for the theta functions. We also present results for the chiral 2-form potential in six dimensions which can be readily extended to $4k+2$ dimensions.}
\end{minipage}
\end{center}
\renewcommand{\thefootnote}{\arabic{footnote}}
\setcounter{footnote}{0}
\newpage 
\tableofcontents

\section{Introduction}

Self-dual  $2k+1$-forms in $4k+2$ dimensions are ubiquitous in physics. They enter the description of the two-dimensional Quantum Hall effect, appear in the spectrum of string theory as the  Ramond-Ramond 5-form flux in ten-dimensional Type-IIB supergravity, and M-theory in the low-energy effective field theory of a single M5-brane. There are, however, well-known difficulties in writing down Poincaré-invariant Lagrangians for self-dual fields and many alternatives that side-step this issue have been put forward over the years \cite{Siegel:1983es,Floreanini:1987as,Imbimbo:1987yt,Bernstein:1988zd,Henneaux:1988gg,McClain:1990sx,Pasti:1995tn,Pasti:1995us,Perry:1996mk,Pasti:1996vs,Pasti:1997gx,Aganagic:1997zq,Witten:1997sc,Witten:1999vg,Belov:2006jd,Mkrtchyan:2019opf}.

More recently, Sen proposed a novel, string-field-theory inspired action for self-dual forms that maintains manifest Lorentz invariance, \cite{Sen:2015nph,Sen:2019qit}. His construction combines two fields: a $2k$-form appearing with wrong-sign kinetic term, and a $2k+1$-form that is self-dual with respect to the flat reference metric $\eta$. The physical spacetime metric $g$ enters the Lagrangian in a non-standard way and as a result diffeomorphism invariance takes a novel form. The action is quadratic in these two fields but they are non-trivially coupled. However, in the Hamiltonian description, the wrong-sign unphysical degrees of freedom explicitly decouple from the physical ones. Furthermore only the physical degrees of freedom are sensitive to the physical metric and external sources. One is then left with a positive-definite Hamiltonian, which correctly captures the physics of self-dual forms with respect to the physical  spacetime metric $g$. This action was supersymmetrised for the case of the self-dual 3-form in the abelian six-dimensional (2,0) theory in \cite{Lambert:2019diy} (wherein an interacting non-abelian generalisation was also constructed), and certain geometric aspects of the formalism were elucidated in \cite{Andriolo:2020ykk}. 

Turning to the quantum theory, and given the complications of finding action principles for self-dual fields, a popular approach for obtaining the chiral form partition function \textit{via} a path-integral computation is so-called ``holomorphic factorisation": One starts with the action for the non-chiral version of the field, evaluates the corresponding path integral in the Wick-rotated theory, observes that the result essentially factorises and reads off the chiral part \cite{Alvarez-Gaume:1986nqf,Alvarez-Gaume:1987wwg,Henningson:1999dm}. In a more rigorous approach,\footnote{For a Hamiltonian version see also \cite{Freed:2006ya,Freed:2006yc}.} the partition function of the self-dual field is determined via a Chern--Simons ``holographic" description in $4k+3$ dimensions \cite{Witten:1988hf,Witten:1996hc,Belov:2006jd} that  takes into account the topological aspects associated with the quantisation of the self-dual field. For example, it is well established that such a partition function is not uniquely defined without the choice of a background spin structure,
or equivalently the choice of a holomorphic line bundle, of which the partition function is a section. Within the holographic approach of \cite{Belov:2006jd}, one is naturally forced to make such a choice when imposing the Gauss law constraint on Chern--Simons theory in one dimension higher.

Since the information about the background spin structure should in principle be encoded within the action in the form of topological terms \cite{Alvarez-Gaume:1986nqf, Alvarez-Gaume:1987wwg}, one could naturally ask whether candidate actions for chiral forms in $4k+2$ dimensions can reproduce the chiral partition function {\it via} a path integral formulation, without resorting to the Chern--Simons description. In this paper we use the action of \cite{Sen:2015nph,Sen:2019qit} as a starting point for precisely such a calculation for the chiral boson on the torus.\footnote{One can attempt to carry out a path integral approach by introducing an infinite number of auxiliary fields \cite{McClain:1990sx,Devecchi:1996cp}. For another construction see \cite{Chen:2013gca}.} An immediate obstacle pertains to Wick rotating the action to Euclidean signature, as the standard analytic continuation to imaginary time leads to a path integral that does not converge because of the wrong-sign nature of one of the fields. Moreover, one may wonder how to impose a self-duality constraint when the signature changes and the Hodge star no longer squares to the identity. Here, however, we employ an alternative prescription for Wick rotating {\it via} a complex deformation of the physical spacetime metric, as suggested by Visser \cite{Visser:2017atf}. It is a happy coincidence that the non-standard coupling of the fields in the Sen action to the background metric is precisely such that the resulting path integral is convergent. Furthermore, as this alternative Wick rotation does not modify the reference metric, the self-duality constraint is unaffected and the physical degrees of freedom of the system are explicitly preserved.\footnote{Note that the idea of Wick rotating a theory {\it via} complex deformations of the spacetime metric has very recently re-emerged in the context of determining which class of complex geometries can a generic quantum field theory be consistently coupled to, and the possible implications of such a procedure for quantum gravity \cite{Kontsevich:2021dmb,Witten:2021nzp,Visser:2021ucg}.}

This allows us to proceed with the evaluation of the path integral. Note that in our calculation both the physical and unphysical modes contribute to the partition function. In this way, through a collection of delicate but ordinary manipulations, we recover the standard expression for the chiral-boson partition function in the form of the ratio $\theta (  T)/\eta(T)$, where $T$ is related to the complex structure of the torus by an $SL(2,{\mathbb Z})$ transformation.\footnote{We appreciate that the moniker ``partition function'' could be taken as an abuse of language. What we mean is the path integral evaluated on a spacetime that is a Euclidean torus.} It is interesting to point out that,  when the radius-squared of the chiral boson is rational (corresponding to a rational conformal field theory), the resulting partition function is an extended  $\widehat{\mathfrak{u}}(1)$ character as expected \cite{Moore:1988qv,Moore:1989yh,Elitzur:1989nr}. In our calculation the choice of spin structure corresponds to introducing topological terms ({\it i.e.} terms that do not affect the equations of motion) to the Sen action along with a  change in the boundary conditions for the fields. We make appropriate choices for such terms, leading to more general theta characteristics in the $\theta (  T )/\eta(T)$ result, once again as expected for the widely studied chiral boson. 

The extension of these results to $4k+2$ dimensional theories is of great interest and we initiate this study by evaluating the path integral of the six-dimensional version of the Sen action on the six-torus, {\it i.e.} $k=1$, including an additional topological term. One recovers once again appropriate generalisations of the two-dimensional answer, involving higher theta functions with general characteristics. Our results here can be readily extended to more general values of $k$. 

The rest of this paper is organised as follows. We begin in Section~\ref{background} with a summary of the salient features of the Sen action \cite{Sen:2015nph,Sen:2019qit}, as well as the Wick-rotation prescription of \cite{Visser:2017atf}. We continue in Section~\ref{2DPI} with the implementation of the Wick rotation and evaluation of the path integral for the chiral boson on the torus. We calculate the oscillator and winding-mode contributions to obtain the $\theta (  T )/\eta(T)$ result, with the more general theta characteristics following suit after adding appropriate topological terms to the action and modifying the boundary conditions. In Section~\ref{6DPI} we sketch the corresponding setup for the chiral 2-form potential in six dimensions and calculate the oscillator and winding-mode contributions to the path integral.  We conclude with some closing comments and open questions in Section~\ref{conclusions}.

\section{Background}\label{background}

We begin our discussion by providing some background material that will form the starting point of our calculation. 

\subsection{Sen Action for Chiral Forms}

In \cite{Sen:2015nph,Sen:2019qit} Sen put forward the following action as a candidate for capturing the physics of chiral forms on a $4k+2$-dimensional spacetime with general metric:
\begin{align}\label{Sen}
   S   = \frac{1}{(2\pi)^{4k+1}}\int\left(\frac12 d P\wedge\star_\eta d P -2Q\wedge d P +   Q\wedge { \widetilde{ {\cal M}}}(Q)  \right)\;,
\end{align}
where $P$ is a $2k$ form with the wrong-sign kinetic term and $Q = \star_\eta Q$ is a self-dual $2k+1$-form. Note that even though the spacetime metric $g$ is non-trivial, the Hodge star is evaluated with respect to the Minkowski metric $\eta$. Instead, the non-triviality of the background is encoded in the term $\widetilde{ {\cal M}}(Q)$, where $\widetilde{ {\cal M}}$ is a map from $\star_\eta$-self-dual forms to $\star_\eta$-anti-self-dual forms such that
\begin{align}
    \mathfrak{m}(\omega):=\omega-\widetilde{ {\cal M}}(\omega)\ ,
\end{align}
is self-dual with respect to $\star_g$.
The action \eqref{Sen} is invariant under infinitesimal diffeomorphisms, $x^\mu \to x^\mu + \xi^\mu(x)$, up to the boundary term 
\begin{align}\label{DiffeoBoundary}
 \delta_\xi S =   \frac{1}{(2\pi)^{4k+1}}\int d\left(\delta_\xi P\wedge dP\right)\ ,
\end{align}
through the non-standard transformations
\begin{align}\label{DiffeoTranf}
    \delta_{\xi} Q & =-\left(\frac{1+\star_{\eta}}{2}\right) d \delta_{\xi} P\cr
    \delta_{\xi} P &=i_{\xi} (Q - \widetilde{ {\cal M}}(Q))\cr
    \delta_{\xi} \widetilde{ {\cal M}}(Q)& =\left(\frac{1-\star_{\eta}}{2}\right) \mathfrak{m}^{-1}(\xi(\mathfrak{m}(Q)))\;,
\end{align}
where
\begin{align}
\label{beforeFluxes}
    \xi(\omega):=\frac{1}{(p-1) !} \nabla_{\mu} \xi^{\lambda} \omega_{\lambda \mu_{1} \ldots \mu_{p-1}} d x^{\mu} \wedge d x^{\mu_{1}} \wedge \ldots \wedge d x^{\mu_{p-1}}\ ,
\end{align}
for any $\omega=\frac{1}{p!}\omega_{\nu_1...\nu_p}dx^{\nu_1}\wedge...\wedge dx^{\nu_p}$.\footnote{Our conventions are those of \cite{Andriolo:2020ykk} where the $\star_g$ operation in D spacetime dimensions is given by
\begin{align*}
\varepsilon_{012...(\mathrm{D}-1)\mathrm{D}}&=-1\\
 \star_g \omega&=\sqrt{-g}\frac{1}{p!}\frac{1}{(\mathrm{D}-p)!} \varepsilon_{\mu_1...\mu_{\mathrm{D}-p}\nu_1...\nu_p}g^{\nu_1\rho_1}...g^{\nu_p\rho_p}  \omega_{\rho_1...\rho_p}dx^{\mu_1}\wedge...\wedge dx^{\nu_{\mathrm{D}-p}}
\end{align*}
and analogously for $\star_\eta$.} 

A closed-form expression for $\widetilde{ {\cal M}}(Q)$ was proposed in \cite{Andriolo:2020ykk} under the assumption that it is symmetric; $Q_{1} \wedge \widetilde{\mathcal{M}}\left(Q_{2}\right)=Q_{2} \wedge \widetilde{\mathcal{M}}\left(Q_{1}\right)$, and $\star_\eta$-anti-self-dual; as only these combinations appear in \eqref{Sen}.\footnote{For an alternative construction see \cite{Vanichchapongjaroen:2020wza}.} This construction considers a basis of $\star_\eta$-(anti)self-dual forms
\begin{align}\label{selfdualbasis}
    \omega^A_+, \     \omega_{-A},\qquad \mathrm{for} \qquad A = 1,\ldots,\frac{1}{2}\binom{4k+2}{2k+1}\ ,
\end{align}
on which
\begin{align}
    \widetilde{ {\cal M}}\left(\omega_{-A}\right)=0\;, \quad\widetilde{ {\cal M}}\left(\omega_{+}^{A}\right)=\widetilde{ {\cal M}}^{A B} \omega_{-B}\;,\quad \mathrm{with}\quad \widetilde{ {\cal M}}^{AB} = \widetilde{ {\cal M}}^{BA}\;.
\end{align}
In terms of this basis, a $\star_g$-self-dual $2k+1$-form can be expanded as
\begin{align}
    \varphi^{A}=\mathcal{N}^{A}{ }_{B} \omega_{+}^{B}+\mathcal{K}^{A B} \omega_{-B}\ .
\end{align}
By demanding that $\omega_{+}^{A}-\widetilde{ {\cal M}}^{A B} \omega_{-B}$ is indeed $\star_g$-self-dual one determines that
\begin{align}\label{Mform}
\widetilde{ {\cal M}}^{A B}=-\left(\mathcal{N}^{-1}\right)^{A}{ }_{C} \mathcal{K}^{C B}\;.
\end{align}
The following combinations of Lagrangian fields
\begin{align}\label{sandg}
Q_{(s)} &:=Q+\left(\frac{1+\star_{\eta}}{2}\right) d P \cr
Q_{(g)} &:=Q-\widetilde{ {\cal M}}(Q)\;,
\end{align}
correspond on-shell to a singlet and a standard self-dual 3-form under diffeomorphisms, hence they are respectively identified with the unphysical and physical chiral degrees of freedom of the system. In the Hamiltonian formulation of the theory, the degrees of freedom $Q_{(s)}$ and $Q_{(g)}$  decouple from each other and one isolates the physical chiral form, satisfying $Q_{(g)}=\star_g Q_{(g)}$  \cite{Sen:2019qit,Andriolo:2020ykk}. 

It is straightforward to introduce sources to \eqref{Sen} by extending the action to\footnote{The source $J$ is a standard $(k+1)$-form, which transforms as usual under diffeomorphisms. To see how \eqref{DiffeoTranf} and \eqref{sandg} get modified when $J\not=0$, see \cite{Andriolo:2020ykk}.}
\begin{align}\label{sensources}
    S  =\frac{1}{(2\pi)^{4k+1}} \int \left(\frac12 d P\wedge\star_\eta d P -2Q\wedge d P +   (Q+J)\wedge \widetilde{ {\cal M}}(Q+J) + 2Q\wedge J-\frac12 J\wedge\star_{\eta} J \right)\;.
\end{align}
This action enjoys a standard gauge symmetry through shifting $P$ by an exact $2k$-form. It is interesting to also observe that under the following wider class of transformations 
\begin{align}\label{Lsym}
P &\to P+\Lambda \cr
J &\to J + d \Lambda \cr
Q &\to Q-\left(\frac{1+\star_{\eta}}{2}\right) d \Lambda\ ,
\end{align}
the action \eqref{sensources} 
changes by
\begin{align}\label{SchangeSources}
S\to S + \frac{1}{(2\pi)^{4k+1}}\int d\Lambda \wedge ( dP-J)\ .
\end{align}
It should be appreciated that the last term in \eqref{sensources} does not contribute to the equations of motion for $P, Q$. Nevertheless, one needs to add it to the action so as to achieve \eqref{SchangeSources}, which we will see  plays an important role in the interpretation of the partition function of the chiral boson.

Despite the non-standard coupling of the fields to the background metric, the action \eqref{sensources} passes various consistency checks for $k=0,1$, when compactified on a torus and respectively on a circle, K3 and a non-compact Riemann surface \cite{Sen:2019qit,Andriolo:2020ykk}.

\subsection{Wick-Rotation via Metric Deformation}
\label{sec:visser}

When evaluating the path integral, one usually passes to the Euclidean version of the theory to ensure convergence. However, applying the standard Wick-rotation procedure of analytically continuing to imaginary time for \eqref{sensources} leads to a non-convergent answer because of a wrong sign for the $P$ kinetic term along with the $dP\wedge Q$ mixing term. A new prescription is also needed since, due to the self-duality constraint, one would like to Wick rotate the theory without altering the degrees of freedom of the system.

We will employ the proposal of Visser \cite{Visser:2017atf}, where instead of analytically continuing to imaginary time, one performs a complex deformation of the Lorentzian spacetime metric $g$, and not the coordinates, {\it via} 
\begin{equation}\label{visser}
(g_\epsilon)_{\mu\nu}:=g_{\mu\nu}+i\epsilon\frac{V_\mu V_\nu}{V^\lambda V_\lambda}\;.
\end{equation}
Here, $\epsilon$ is the deformation parameter, $V^\mu$ is an arbitrary, nowhere-vanishing timelike vector field (which is guaranteed to exist  because the spacetime has a global Lorentzian signature) and $V_\mu$ is the associated co-vector, \emph{i.e.} $V_\mu:=g_{\mu\nu}V^\nu$. For the simple case of flat space and a constant vector $V_\mu = (-1,0,\ldots,0)$, \eqref{visser} results in
\begin{align}
    g_{\epsilon}=\mathrm{diag}(-1-i \epsilon,+1,\ldots,+1) \;,
\end{align}
which recovers the standard Minkowski metric for $\epsilon = 0$ and the Euclidean metric for  $\epsilon = 2 i $. In fact this case is completely equivalent to the standard Wick rotation via analytic continuation to imaginary time; see \cite{Visser:2017atf}. Following this, the prescription for analytically continuing to Euclidean signature for arbitrary metrics consists of taking \eqref{visser} and setting $\epsilon = 2 i$.\footnote{One cannot make this simply a real deformation by {\em e.g.} setting $i \epsilon = \lambda$, and taking $\lambda$ from $0 \rightarrow -2$, as the metric would become singular at $\lambda = -1$ or $\epsilon = i$. The complex deformation allows us to go around this point in the $\epsilon$-plane.} Note that the resulting Euclidean metric is in general not unique but depends on the choice of constant timelike vector $V^\mu$. 

Although this recipe was initially put forward to produce Euclidean metrics that are compatible with the existence of a Lorentzian metric \cite{Visser:2017atf}, it is particularly apt in our case where the physical metric dependence of \eqref{sensources} comes entirely through $\widetilde{\mathcal M}$. We stress that the reference metric $\eta$ is left untouched by this Wick rotation, bringing in two advantages. First, we can keep the original constraint $Q=\star_\eta Q$ even in the Wick-rotated theory, and this guarantees that the latter describes the same number of degrees of freedom as the Lorentzian one. Second, the Wick rotation \eqref{visser} makes the path integral convergent. This happens because the wrong-sign term $dP\wedge \star_\eta dP$ remains a purely oscillatory contribution, and this allows us to immediately single out a holomorphic partition function.

 It is noteworthy that very recently the idea of Wick rotating a theory via complex deformations of the spacetime metric has re-appeared in an attempt to replace the standard axioms of QFT with the requirement that they be consistently coupled to complex metrics \cite{Kontsevich:2021dmb}. Applications of this proposal to quantum gravity were considered in \cite{Witten:2021nzp}. In these works, a complex metric is allowed, under the condition that it leads to a convergent path integral. As will become clear such constraints arise naturally for our metric.

\section{The Path Integral for the Two-dimensional Chiral Boson}\label{2DPI}

In this section we will write down a well defined path integral for the two-dimensional chiral boson on ${\mathbb T}^2$, starting from the action of \cite{Sen:2019qit,Andriolo:2020ykk} and analytically continuing to Euclidean signature à la \cite{Visser:2017atf}. We will then evaluate it to directly obtain the chiral boson partition  function with  particular characteristics for the theta function. More general theta-characteristics will be introduced by including boundary terms in the action, and adjusting the periodicities of the scalar field. In this section we will denote the 0-form $P$ by $\phi$ and the 1-form $Q$ by $H$ to keep closer contact with the literature.

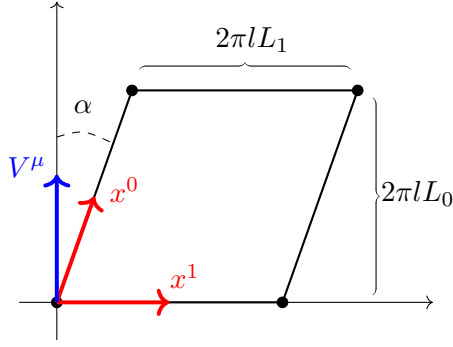
\begin{figure}
    \centering
\begin{tikzpicture}[scale=1, every node/.style={transform shape},baseline={([yshift=-.5ex]current bounding box.center)},node/.style={anchor=base,
    circle,fill=black!25,minimum size=18pt,inner sep=2pt}]]
\draw (0,0);
\draw[->] (-0.5,0) to (5,0);
\draw[->] (0,-0.5) to (0,4);
\filldraw (0,0) circle (2pt);
\filldraw (3,0) circle (2pt);
\filldraw (1,2.82) circle (2pt);
\filldraw (4,2.82) circle (2pt);
\draw[thick] (0,0) to (3,0);
\draw[thick] (0,0) to (1,2.82);
\draw[thick]  (1,2.82) to (4,2.82);
\draw[thick] (4,2.82) to (3,0);
\draw[dashed] (0,2.2) .. controls(0.35,2.3).. (0.75,2.115) ;
\draw (0.35,2.6) node {$\alpha$};
\draw [decorate,
    decoration = {calligraphic brace}] (4.2,2.72) --  (4.2,0.1);
\draw (4.8,1.45) node {$2\pi l L_0$};    
\draw [decorate,
    decoration = {calligraphic brace}] (1.1,3.1) --  (3.9,3.1);
\draw (2.6,3.5) node {$2\pi l L_1$};  
\draw[red,line width=1.5pt,->] (0,0) to (0.5,1.41);
\draw[red,line width=1.5pt,->] (0,0) to (1.5,0);
\draw (0.9,1.5) node {\textcolor{red}{$x^0$}};
\draw (1.7,0.35) node {\textcolor{red}{$x^1$}};
\draw[blue,line width=1.5pt,->] (0,0) to (0,1.7);
\draw (-0.4,1.8) node {\textcolor{blue}{$V^\mu$}};
\end{tikzpicture} 
    \caption{The flat torus parametrised by the coordinates $x^0,x^1$, in red. In blue, the time-like vector that will be used to perform the Wick rotation.}
    \label{fig:torus}
\end{figure}

To start we define our path integral on the torus of Figure~\ref{fig:torus} using coordinates $(x^0,x^1)$, subject to the identifications
\begin{align}
x^0 \cong x^0 + 2 \pi l\;, \quad x^1 \cong x^1 + 2\pi l \;,
\end{align}
where $l$ is an arbitrary length scale. The metric is then dimensionless and in these coordinates reads 
\begin{equation}
g_{\mu\nu}= \left(\begin{matrix}
 -L_0^2+L_0^2\tan^2 \alpha & L_1L_0 \tan \alpha \\L_1L_0 \tan \alpha  &L_1^2
\end{matrix}
\right)\;.
\end{equation}
 Note that  the choice of constant timelike vector $V^\mu$ needed to implement the Wick rotation  in \eqref{visser} is not unique. We find it natural to use the timelike vector $V^\mu := (1/L_0,-\tan\alpha/L_1)$, see Figure~\ref{fig:torus}. With this choice, the deformed metric becomes
\begin{equation}
(g_{\epsilon})_{\mu\nu}= \left(\begin{matrix}
 -(1+i\epsilon)L_0^2+L_0^2\tan^2 \alpha & L_1L_0\tan \alpha \\L_1L_0\tan \alpha  &L_1^2
\end{matrix}
\right)\;,
\label{Visser}
\end{equation}
 which leads to the usual flat metric on the flat Euclidean torus for $\epsilon = 2 i $.  Note that the determinant of $g_{\epsilon}$ 
is proportional to $ (1+i\epsilon)$, so the analytic continuation must be performed by
avoiding $\epsilon=+i$.\footnote{We also avoid $\epsilon\in i\mathbb{R}_{<0}$, as
we have implicitly placed the branch cut of
$\sqrt{-g_{\epsilon}}=L_1L_0\sqrt{1+i\epsilon}$ there.} From now on we will set $\epsilon = 2i$. 

We next define the 1-forms
\begin{align}
\omega_+&=dx^0-dx^1\nonumber\\
\omega_-&=dx^0+dx^1\ ,
\end{align}
satisfying
\begin{align}
\star_\eta \omega_{\pm}&=\pm\omega_\pm\nonumber\\
\omega_+\wedge \omega_-&=2dx^0 \wedge dx^1\;.
\end{align}
Following the discussion around Eqs \eqref{selfdualbasis}-\eqref{Mform} one  arrives at the expression:\footnote{So as to arrive at a convergent path integral we pick the negative branch for the square root on the complex plane ($\sqrt{-1}=-i$).}
\begin{align}\label{M11}
 \widetilde{ {\cal M}}^{11} =\frac{\frac{L_1}{L_0}+\tan \alpha+i}{\frac{L_1}{L_0}-\tan\alpha - i }\;.
\end{align}
By further introducing the complex structure
\begin{equation}
\tau=\frac{L_0}{L_1}(\tan \alpha+i)\ ,
\end{equation}
we can rewrite \eqref{M11} more simply as
\begin{align}
\widetilde{ {\cal M}}^{11} =-\frac{\tau+1}{{\tau}-1} =: \mathcal M \;.
\label{MwickRotated}
\end{align}
Remarkably, $\mathcal M$ is a meromorphic function of $\tau$, and since $\text{Im}(\tau)$ is strictly positive, so is $\text{Im}({\mathcal M})$.

We can now  expand
\begin{align}
H &= H^+ \omega_+ \nonumber \\
 J &= J^{+} \omega_{+}+J^{-} \omega_{-} \;,
\end{align}
in terms of which the action (\ref{sensources}) becomes
\begin{align}
S =\frac{1}{2\pi} \int_{\mathbb T^2} d^2x\Big(-\frac{1}{2} (\partial_0 \phi)^2 + \frac{1}{2} (\partial_1 \phi)^2 &- 2 H^+ ( \partial_0 \phi + \partial_1 \phi) \cr
& + 2 {\mathcal M} (H^+ + J^{+})^2  + 4H^{+} J^{-} + 2 J^{+}J^{-} \Big)\ .
\label{eq:Sfull}
\end{align}

\subsection{The Dirac Path Integral Prescription}\label{sec:dirac}

Since we are dealing with a constrained system, we  would now like to evaluate the Dirac path integral for the action \eqref{eq:Sfull}. 
Due to $\star_\eta$-self-duality $H_0=H^+=-H_1$ and we can identify $\phi$ and $H_1$ as field variables with canonical conjugate momenta 
\begin{align}
\Pi^\phi&:=\frac{\delta S}{\delta \partial_0\phi}=\frac1{2\pi}\left(-\partial_0\phi+2H_1\right)\;, \qquad \Pi^{H_1}:=\frac{\delta S}{\delta \partial_0 H_1}=0\;.
\end{align}
The Hamiltonian density is thus
\begin{align}
\mathcal{H}=\frac1{2\pi}\left[-2 (\pi\Pi^\phi-H_1)^2-\frac12 (\partial_1 \phi)^2-2H_1\partial_1\phi-2(J^+-H_1)^2\mathcal{M}+4H_1J^--2J^+J^-\right]\ ,
\end{align}
and the constraint surface is defined by\footnote{Given two functionals $F,G$ of the fields $\phi^i$ and their conjugate momenta $\Pi^i$, we denote with $\{F,G\}_{x^0}$ their Poisson bracket at equal time, {\it i.e.}
\begin{align}
    \{F,G\}_{x^0}=\int dx^1 \frac{\delta F}{\delta \phi^i(x^0,x^1) } \frac{\delta G}{\delta \Pi^i(x^0,x^1)}- \frac{\delta G}{\delta \phi^i(x^0,x^1) }\frac{\delta F}{\delta \Pi^i(x^0,x^1)}\;.
\end{align}
}   
\begin{align}
\chi_1(x^0,x^1)& :=\Pi^{H_1}(x^0,x^1)    \cr
\chi_2(x^0,x^1) &:=\left\{\chi_1(x^0,x^1),\int_{S^1}dy^1\mathcal H(x^0,y^1)\right\}_{x^0}\cr
&=-\frac2\pi\left[\pi\Pi^\phi-H_1-\frac12\partial_1 \phi+(J^+-H_1)\mathcal{M}+J^-\right](x^0,x^1)\;,
\end{align}
 which are primary and secondary constraints respectively. The constraints $\chi_1$ and $\chi_2$ form a pair of second-class constraints,\footnote{There are no first-class constraints for this system. This fact is particular to two dimensions since, when working in $4k+2$ dimensions with $k>0$, the local gauge redundancy of $P$ should be taken into account. We will come back to this point in Section~\ref{6DPI}.} since their Poisson bracket reads 
\begin{align}
\{\chi_1(x^0,x^1),\chi_2( x^0, y^1)\}_{x^0}=-\frac2\pi(1+\mathcal M)\delta(x^1-y^1)\;.
\end{align}
The Dirac path integral is then given by  \cite{Senjanovic:1976br,Henneaux:1992ig}
\begin{align}
Z[J^\pm,\mathcal M ]=&\int[\mathcal D\phi\mathcal D \Pi^\phi\mathcal D H_1\mathcal D \Pi^{H_1}]\;\delta\left(\pi\Pi^\phi-H_1-\frac12\partial_1 \phi+(J^+-H_1)\mathcal{M}+J^-\right)\cr
&\qquad\qquad \times\delta\left(\Pi^{H_1}\right)\sqrt{\det\{\chi_i,\chi_j\}_{x^0}}\times e^{i\int_{\mathbb T^2}d^2 x\left(\partial_0\phi \Pi^\phi+\partial_0H_1\Pi^{H_1}-\mathcal H\right)}\;.
\end{align}
Up to an overall constant factor
\begin{align}
\delta\Big(\pi\Pi^\phi-H_1-\frac12\partial_1 \phi &+(J^+-H_1)\mathcal{M}+J^-\Big)
\sqrt{\det\{\chi_i,\chi_j\}_{x^0}}\propto\cr
&\propto\delta\left(H_1-\frac{1}{1+\mathcal M}\left(\pi\Pi^\phi-\frac12\partial_1 \phi+J^+\mathcal M+J^-\right)\right)\ ,
\end{align}
and the functional integration over $\Pi^{H_1}$ and $H_1$ can thus immediately be performed to get
\begin{align}
\label{Crossroad}
Z[J^\pm,\mathcal M ]=\int[\mathcal D\phi\mathcal D\Pi^\phi] e^{i\int_{\mathbb T^2}d^2 x\left[\partial_0\phi \Pi^\phi-\mathcal H\right]}\Bigg|_{H_1\to\frac{1}{1+\mathcal M}\left(\pi\Pi^\phi-\frac12\partial_1 \phi+J^+\mathcal M+J^-\right)}\;.
\end{align}

In this framework, the first thing to do is to exploit the translational invariance over the space of $\Pi^\phi$, to factorise the path integral \eqref{Crossroad} into
\begin{align}
Z[J^\pm,\mathcal M]= \mathcal{W}[J^\pm, {\mathcal M}] \int [\mathcal D\Pi^\phi] e^{i\pi\frac{\mathcal M}{\mathcal M+1}\int_{\mathbb T^2}d^2 x(\Pi^\phi)^2}  \int[\mathcal D\phi]e^{i S_{\text{eff}}[\phi,J^\pm]}\;,
\label{AlmostDirac}
\end{align}
where we have defined the overall field-independent function of the sources
\begin{align}\label{anomalous}
 \mathcal{W}[J^\pm,{\mathcal M}]    :=e^{-\frac{i}{\pi} \int_{\mathbb T^2} d^2x (J^+J^-+{\mathcal M}^{-1}(J^-)^2)} \ ,
\end{align}
and
\begin{align}
  S_{\text{eff}}[\phi,J^\pm]
&:=\frac{1}{2\pi}\int_{\mathbb T^2}d^2 x \Big(-\frac12(\partial_0\phi)^2+\frac12(\partial_1\phi)^2-\frac{1}{2\mathcal M}\left(\partial_0\phi+\partial_1\phi\right)^2\cr 
&\qquad\qquad\qquad\qquad\qquad+\frac{2}{\mathcal M}(\partial_0\phi+\partial_1\phi)(\mathcal M J^++J^-)\Big)
\label{Sen'sEffective}\;.
\end{align}

Note that in order to arrive at \eqref{AlmostDirac}, we just used the fact that $\eta$ is the (reference) Minkowski metric; no assumptions were made about the physical metric $g$, the data of which is contained inside $\mathcal M$. 
Note also that once we Wick rotate the physical metric as in \eqref{Visser}, both functional integrations over $\Pi^\phi$ and $\phi$ yield convergent Gaussian integrals, see \eqref{MwickRotated}. 

To proceed with the evaluation, we expand $\Pi^\phi$ in terms of an $L^2(\mathbb{T}^2)$  basis, {\it i.e.}
\begin{align}
\Pi^\phi(x^0,x^1)&=\sum_{n_0,n_1\in\mathbb{Z}^2}\Pi^\phi_{n_0,n_1}e^{-i\frac{x^0}{l}n_0}e^{-i\frac{x^1}{l}n_1}\ ,
    \end{align} 
with $\Pi^\phi_{-n_0,-n_1}=(\Pi^\phi_{n_0,n_1})^*$ such that 
\begin{align}
\int [\mathcal D\Pi^\phi]&\sim \prod_{(n_0,n_1)\in \mathbb{Z}^2}\int d\Pi^\phi_{n_0,n_1}\quad.
\end{align}
Performing a complex integral  over all $\Pi^\phi_{n_0,n_1} $ with $(n_0,n_1)\in {\mathbb Z}^2$ would double count the independent fields. Therefore we need to restrict to a domain 
$U\subset\mathbb{Z}^2$ that has the property that $U\cap (-U)=\emptyset$. Furthermore $\Pi^\phi_{0,0}$ must be treated separately as it is real so $(0,0)\notin U$. But when combined we need $U\cup (-U)\cup(0,0)={\mathbb Z}^2$. To this end we find it helpful to define
\begin{align}
\label{Uset}
U:=\{(n_0,n_1)\in\mathbb{Z}^2| n_0+n_1> 0\}\cup\{(n_0,-n_0)| n_0\in \mathbb N \}\ .
\end{align}
Pictorially we can think of this as the set of parallel diagonal lines $(n_0+p,-n_0)$ with $n_0\in \mathbb Z$ for a fixed $p\in\mathbb N$  along with the half-line $(n_0,-n_0)$ with $n_0\in \mathbb N$. Then  we can easily rewrite 
\begin{align}
\label{PathIntegralMeasure}
\prod_{(n_0,n_1)\in \mathbb{Z}^2}\int d\Pi^\phi_{n_0,n_1} &=\int_{\mathbb{R}}d\Pi^\phi_{0,0} \prod_{(n_0,n_1)\in U}\int_{\mathbb{C}}d\Pi^\phi_{n_0,n_1}d(\Pi^\phi_{n_0,n_1})^*\cr
&= \int_{\mathbb{R}}d\Pi^\phi_{0,0}  \prod_{(n_0,n_1)\in U}\left(2\int_{\mathbb{R}^2}d\mathrm {Re}[\Pi^\phi_{n_0,n_1}]d\mathrm{Im}[\Pi^\phi_{n_0,n_1}]\right)\ ,
\end{align}
and hence find for the $\Pi^\phi$ functional integral
\begin{align}
\int [\mathcal D\Pi^\phi] e^{i\pi\frac{\mathcal M}{\mathcal M+1}\int_{\mathbb T^2}d^2x(\Pi^\phi)^2}&=\sqrt{\frac{i}{(2\pi l)^2}\frac{\mathcal M+1}{\mathcal M}} \left(2\frac{i}{(2\pi l)^2}\frac{\mathcal M+1}{\mathcal M}\right)^{|U|}\;,
\end{align}
where  $|U|$ is the cardinality of $U$. 

To compute $|U|$ we note that the half-line $(n_0,-n_0)$, $n_0\in \mathbb N$ contributes $\sum_{n_0=1}^\infty 1=\zeta(0)=-1/2$ to $|U|$. On the other hand, each complete line $(n_0+p,-n_0)$ consists of two half-lines plus a point and hence contributes $2\zeta(0)+1=0$ to $|U|$. Thus we simply find
\begin{align}
\label{Uregularized}
	|U|=-1/2\ ,
\end{align}
and so
\begin{align}\label{regularised}
\int [\mathcal D\Pi^\phi] e^{i\pi\frac{\mathcal M}{\mathcal M+1}\int_{\mathbb T^2}d^2x(\Pi^\phi)^2}\sim 1\;.
\end{align}
From here onwards we will use $\sim$ to denote equality of the partition function up to an irrelevant---although possibly infinite---constant.

All in all, we have reduced the functional integral \eqref{AlmostDirac} to the evaluation of  the Feynman path integral for the effective action  \eqref{Sen'sEffective}:
\begin{align}
Z[J^\pm,\mathcal M]&\sim  \mathcal{W}[J^\pm, {\mathcal M}] \int[\mathcal D\phi]e^{i S_{\text eff}[\phi, J^\pm]}\;.
\label{EffectiveActionPathIntegral}
\end{align}

This result, reached using the Dirac path integral  and employing the regularisation \eqref{Uregularized}, can in fact be reproduced  by considering the Feynman path integral for the original action. To see this  we complete the square on $H^+$ in \eqref{eq:Sfull} by introducing
\begin{align}
\hat H^+ =  H^+ + \left(J^{+} + \frac{1}{\mathcal M} J^{-}\right) - \frac{1}{2{\mathcal M}} ( \partial_0 \phi + \partial_1 \phi)  \ . 
\end{align}
By construction $H$ is an arbitrary $\star_\eta$-self-dual form and hence $[\mathcal DH^+]=[\mathcal D\hat H^+]$. Thus the Feynman path integral for \eqref{eq:Sfull} factorises into 
\begin{align}
Z =  \mathcal{W}[J^{\pm}, {\mathcal M}] Z_H({\mathcal M})  \int [ \mathcal{D} \phi] e^{i S_{\text{eff}}[\phi,J^\pm]} \ ,
\end{align}
where  
\begin{align}
Z_H({\mathcal M}) & = \int [\mathcal{D} \hat H^+] e^{\frac{i}{\pi} \int_{\mathbb T^2} d^2 x \,  {\mathcal M} (\hat H^+)^2}\;.
\end{align}
Since the imaginary part of ${\mathcal M}$ is positive, the integration over $\hat H^+$ converges and using the same regularisation as for \eqref{regularised}, one recovers an overall constant
\begin{align}\label{Feyn}
    Z_H({\mathcal M}) \sim 1 \;.
\end{align}
 Encouraged by this agreement we will directly employ the Feynman path-integral description in our upcoming discussion, Sections~\ref{general} and \ref{6DPI}.

After this preliminary work, our task now is to evaluate the functional integral \eqref{EffectiveActionPathIntegral}.
We assume that the field $\phi$ is compact with radius $R$, {\it i.e.}
\begin{align}
\phi\cong\phi+2\pi R\ ,
\end{align}
 and it thus admits the following decomposition on the torus
\begin{align}
\phi&= \phi_{\text{w.m.}}+\phi_{\text{osc}}\;.
\end{align}
The $\phi_{\text{w.m.}}$ and $\phi_{\text{osc}}$ respectively encode the winding and oscillatory modes of the field:
\begin{align}
\phi_{\text{w.m.}}&=\frac{R}{l}(m_0x^0+m_1x^1)\;,\qquad\qquad\qquad\qquad\qquad m_0,m_1\in\mathbb{Z}\\
\phi_{\text{osc}}&=\sum^\prime_{n_0,n_1\in\mathbb{Z}^2}\phi_{n_0,n_1}e^{-i\frac{x^0}{l}n_0}e^{-i\frac{x^1}{l}n_1}\;, \qquad\,\,\, \phi_{-n_0,-n_1}=(\phi_{n_0,n_1})^*\quad,
\end{align} 
where the prime symbol on top of the sum denotes that the choice $(n_0,n_1)=(0,0)$ must not be taken into account.

The compact field $\phi$ can admit topologically non-trivial configurations because it appears in the action only through its derivative: the action is still single-valued on the torus even when $\phi$ is not. The same is not true for $J$ which thus admits the following expansion on the torus:
\begin{align}
J=\sum_{n_0,n_1\in\mathbb{Z}^2}J_{n_0,n_1}e^{-i\frac{x^0}{l}n_0}e^{-i\frac{x^1}{l}n_1}\;, \qquad\,\,\, J_{-n_0,-n_1}=(J_{n_0,n_1})^*\quad.
\end{align}
Since the effective action \eqref{Sen'sEffective} is quadratic in $\phi$, $\phi_{\text{w.m.}}$ and $\phi_{\text{osc}}$ decouple. In the following subsections their contributions to the path integral will be determined separately. 

Finally, it is important to note that, in the absence of sources, \eqref{Sen'sEffective} closely resembles the Siegel action \cite{Siegel:1983es}
\begin{align}\label{SiegelAction}
S_{\text{Siegel}}[\phi]&=\frac1{2\pi}\int d^2 x \left(\frac12(\partial_0\phi)^2-\frac12(\partial_1\phi)^2+\lambda\left(\partial_0\phi+\partial_1\phi\right)^2\right)\;,
\end{align}
where $\lambda$ is a Lagrange multiplier imposing the chiral condition on $\phi$. This can also be related to the Floreanini--Jackiw action \cite{Floreanini:1987as,Bernstein:1988zd}. However, one key difference is that ${\mathcal M}$ in \eqref{Sen'sEffective}, unlike $\lambda$ in \eqref{SiegelAction}, is a constant and not a Lagrange multiplier hence the interpretation is different: in our case ${\mathcal M}$ is not a field but is in fact fixed by the complex structure of the spacetime torus.

\subsection{Chiral-boson Partition Function}\label{sec:osc}

At this stage, many of the algebraic manipulations needed for the remainder of this section are standard, see {\it e.g.} \cite{Imbimbo:1987yt,DiFrancesco:1997nk,Chen:2013gca}; we refer the reader interested in the full details to Appendix~\ref{app:details} and only provide a summary of the results.

Evaluating the effective action \eqref{Sen'sEffective} on the oscillator modes we find
\begin{align}
S_{\text{eff}}[\phi_{\text{osc}},J^\pm]&=  \pi \sum_{n_0,n_1}^\prime  \left\{ |\phi_{n_0,n_1}|^2 \left( n_1^2-n_0^2 -  \frac{1}{{\mathcal M}}(n_0+n_1)^2\right)  \right.\nonumber\\
&\left.\qquad\qquad\qquad
+4i l(n_0+n_1) \phi_{n_0,n_1}\big( J^+_{-n_0,-n_1} + \frac{1}{{\mathcal M}} J^-_{-n_0,-n_1} \big) \right\}\ ,
\label{oscillationSmain}
\end{align}
which by letting $n:=n_0+n_1 $ and $m := n_1$ leads to (see Appendix~\ref{app:osc})
\begin{align}
Z_{\text{osc}}&=\prod_{\substack{m\\ n>0}}\, {\frac{i}{  n(m+Tn)}}\;.
\label{zetaregmain}
\end{align}
Here we introduced the important parameter 
\begin{align}
\label{Tparametermain}
    T :=-\frac12 (1+ {\mathcal M}^{-1})= - \frac{1}{1+ \tau}\;,
\end{align}
which in turn can be related to $\tau$ {\it via} an $SL(2, \mathbb{Z})$ transformation.
In Appendix~\ref{sec:appen} we show that the infinite product appearing in \eqref{zetaregmain} can be regularised to
\begin{align}
  \prod_{\substack{m\\ n>0}}\,{\frac{i}{  n(m+Tn)}}\sim \frac{1}{\eta(T)}\
 \end{align}
thus arriving at 
\begin{align}
Z_{\text{osc}} \sim \frac{1}{\eta(T)}\ .
\end{align}

Next we turn to the winding modes, which lead to a non-trivial dependence on the sources of the partition function.
Evaluating the effective action \eqref{Sen'sEffective} on the winding modes returns
\begin{align}
S_{\text{eff}}[\phi_{\text{w.m.}},J^\pm]  &=  \pi \left(   -R^2m_0^2+R^2m_1^2 -  \frac{R^2}{{\mathcal M}}(m_0+m_1)^2 
+2R(m_0+m_1)\mathcal{J}^{(0)} \right)\;,
\end{align}
where we defined a complex structure on the sources by 
\begin{align}
    \mathcal{J} = J^{+} + \frac{1}{\mathcal{M}} J^{-}\ ,
\end{align}
with normalised zero mode defined by
\begin{align}\label{zeromain}
\mathcal{J}^{(0)}  & := \frac{2l}{(2  \pi l)^2  }\int d^2 x \left( J^+ + \frac{1}{\mathcal M} J^- \right)\;.
\end{align}
Following on from the oscillator discussion this leads to (see Appendix~\ref{app:wind})
\begin{align}\label{deltamain}
Z_{\text{w.m.}} &= \sum_{n} \sum_{m}\delta(R^2n-m) e^{2i\pi R^2n^2T +2i\pi R n\mathcal{J}^{(0)}}\;,
\end{align}
where $T$ was defined in \eqref{Tparametermain}. One then finds that 
$Z_{\text{w.m.}}= \delta(0)$ for $R^2$ irrational. Instead, if $R^2$ is rational---that is when $R^2 = {r_1}/{r_2}$ for some coprime integers $r_1, r_2$---one gets a non-vanishing contribution whenever
\begin{align}
n = p r_{2}\qquad\text{and}\qquad m=pr_1 \ ,
\end{align}
for any $p\in\mathbb{Z}$. Hence for $R^2 = {r_1}/{r_2}$ we have
\begin{align}\label{Zwindingmain}
Z_{\text{w.m.}}&= \delta(0) \sum_{p} e^{2i\pi r_1r_2 p^2 T +2i\pi \sqrt{r_1r_2} p\mathcal{J}^{(0)}}\sim\theta \! \begin{bmatrix} \vspace{-20pt} \\ 0  \\ \vspace{-20pt} \\ 0 \\ \vspace{-20pt} \end{bmatrix} \! ( \sqrt{r_1r_2}\mathcal{J}^{(0)} | 2r_1r_2T )\ .
\end{align}
In writing the last step of \eqref{Zwindingmain} we have dropped the $\delta(0)$ as an irrelevant but infinite constant arising from the unphysical chiral modes, originating from the wrong-sign kinetic terms in the Lagrangian (see Appendix \ref{app:wind}).

Combining \eqref{Zwindingmain} with $Z_{\text{osc}}$ and the source-dependent prefactor appearing in \eqref{EffectiveActionPathIntegral}, we have
\begin{align}
Z[J^\pm, T] 
&\sim \mathcal{W}[J^\pm,T] Z_{\text{osc}}Z_{\text{w.m.}}\nonumber\\
&\sim\mathcal{W}[J^\pm,T]   \frac{\theta \! \begin{bmatrix} \vspace{-20pt} \\ 0  \\ \vspace{-20pt} \\ 0 \\ \vspace{-20pt} \end{bmatrix} \! \Big( \sqrt{r_1 r_2} \, \mathcal{J}^{(0)} \, | \,  2r_{1} r_{2}  T \Big)}{\eta(T)} \nonumber\\
&= e^{-\frac{i}{\pi} \int_{\mathbb T^2} d^2x (J^+J^-+{\mathcal M}^{-1}(J^-)^2)} \frac{\theta \! \begin{bmatrix} \vspace{-20pt} \\ 0  \\ \vspace{-20pt} \\ 0 \\ \vspace{-20pt} \end{bmatrix} \! \Big( \sqrt{r_1 r_2} \, \mathcal{J}^{(0)} \, | \,  2r_{1} r_{2}  T \Big)}{\eta(T)}\;.
\label{Zfullmain}
\end{align}
In the absence of sources, this can be interpreted as a $\widehat{\mathfrak{u}}(1)_{r_1 r_2}$ character, as expected for a chiral boson on a rational square radius.\footnote{For the interested reader a useful resource is \cite{DiFrancesco:1997nk}.} Moreover, since \eqref{Zfullmain} is left invariant by $r_1 \leftrightarrow r_2$, we recognise an underlying duality acting  as $R \leftrightarrow 1/R$. This fixes the self-dual radius to be $R^2=r_1/r_2=1$ which, as $r_1$ and $r_2$ are coprime, means setting $r_1 = r_2 = 1$ and we find an $\widehat{\mathfrak{su}}(2)_1$ character. All these results are compatible with the formulation of the chiral boson as the edge mode of abelian Chern--Simons theory in one dimension higher \cite{Moore:1988qv,Moore:1989yh,Elitzur:1989nr}

To summarise, when $R^2$ is rational our computation precisely lands---up to the $SL(2,Z)$ twist encoded in $\tau\to T$---on the expected result for the chiral-boson partition function. For $R^2$ irrational, the theta function collapses to one, and $Z$ is only proportional to $1/\eta(T)$.

\subsection{General Theta-Function Characteristics and  Holomorphic Structure}\label{general}

We now see how the discussion from the previous section can be extended to include general theta-characteristics.
To this end we take
\begin{align}
\phi_{\text{w.m.}} = \frac{R}{l}(m_0+\alpha_0)x^0 + 	\frac{R}{l}(m_1+\alpha_1)x^1\ ,
\end{align}
for some constants $\alpha_0,\alpha_1$. Here $\phi$ is not single-valued over the torus  and now satisfies
\begin{align}
\phi(x^0+2\pi l,x^1) &= \phi(x^0,x^1) + 2\pi R(m_0+\alpha_0)\nonumber\\ 	
\phi(x^0,x^1+2\pi l) &= \phi(x^0,x^1) + 2\pi R(m_1+\alpha_1)\ .
\end{align}
Shifting $\phi$ by a constant is a symmetry of the action, corresponding to a constant choice of $\Lambda$ in (\ref{Lsym}). Thus we can view this identification as an orbifold whose action is different over the two 1-cycles of $\mathbb T^2$. However, we do not want to allow for orbifold actions of the form $\phi \cong \phi + \Lambda$ for any constant $\Lambda$ as that would completely remove all zero modes. So we restrict to $m_0,m_1\in \mathbb Z$. Another way to say this is that we restrict to orbifold actions for which
\begin{align}
	\frac{1}{2\pi}\oint (d\phi - {\cal A})\in R\;{\mathbb Z}\ ,
\end{align} 
where the integral is over any 1-cycle of the torus and 
\begin{align}
{\cal A} = \frac{R}{l}\alpha_0dx^0+\frac{R}{l}\alpha_1 dx^1	\ ,
\end{align}
is a fixed closed 1-form. Note that we can also think of these boundary conditions in terms of  $\Psi= e^{i\phi/R}$:
\begin{align}
\Psi(x^0+2\pi n^0 l,x^1+ 2\pi l n^1)    = e^{2\pi i (n^0\alpha_0+n^1\alpha_1)}\Psi(x^0,x^1) \ .
\end{align}
Thus if we think of $\Psi$ as a dual fermion then ${\cal A}$ encodes the spin structure. It is tempting to interpret ${\cal A}$ as a connection 1-form and $d\phi-{\cal A}$ as a covariant derivative, as in \cite{Witten:1996hc}. However, this interpretation has difficulties in higher dimensions.

Next, we repeat the winding-mode calculation with the more general modings $m_0\to m_0+\alpha_0$, $m_1\to m_1+\alpha_1$, while also adding the following term to the action \eqref{eq:Sfull}:
\begin{align}\label{toptermmain}
S_{\cal A} :=  S - \frac{1}{2\pi}\int_{\mathbb T^2} {\cal A}	\wedge d\phi\ .
\end{align}
This term is a total derivative  and hence does not affect the equations of motion or any of the symmetries, including infinitesimal diffeomorphisms. However, it does give the following contribution on the winding modes if $R^2=r_1/r_2$  (see Appendix~\ref{app:general})
\begin{align}
Z_{\text{w.m.}} & = e^{-\pi i\alpha\beta}\theta \! \begin{bmatrix} \vspace{-20pt} \\ \alpha \\ \vspace{-20pt} \\ \beta \\ \vspace{-20pt} \end{bmatrix} \! ( \sqrt{r_1r_2}\mathcal{J}^{(0)} |2 r_1r_2T )\ ,
\end{align}
where
\begin{align}
\alpha &= (\alpha_0+\alpha_1)/r_2\;,	\qquad \beta = 2r_1\alpha_1\ .
\end{align}

Finally, let us look at the zero-mode contribution to the partition function:
\begin{align}\label{370main}
Z \sim {\cal W}'[J] e^{\frac{i}{2}\pi {\cal J}^{(0)}   ({\cal J}^{(0)}   -\bar{\cal J}^{(0)}  )/(T-\bar T)}\frac{\theta \! \begin{bmatrix} \vspace{-20pt} \\ \alpha  \\ \vspace{-20pt} \\ \beta \\ \vspace{-20pt} \end{bmatrix} \! ( \sqrt{r_1r_2}{\cal J}^{(0)}    | 2r_1r_2T )}{\eta(T)}\ ,
\end{align}
where ${\cal W}'[J]$ only depends on the non-zero-mode sources. This is almost a holomorphic function of ${\cal J}^{(0)}$, which is encoded in the statement that
\begin{align}
\bar {\cal D}Z:=\left(\frac{\partial }{\partial \bar {\cal J}^{(0)}}  +\frac{i\pi }{2}\frac{{\cal J}^{(0)}}{T-\bar T}\right) Z=0  \ .
\end{align}

As a result one sees that under a shift (see Appendix~\ref{app:holo})
\begin{align}\label{IntJmain}
{\cal J}^{(0)} \to   {\cal J}^{(0)}  + \frac{m}{\sqrt{r_1r_2}} + 2n\sqrt{r_1r_2}T \ ,
\end{align}
the overall change to $Z$ is simply
\begin{align} \label{phasemain}
    Z \to e^{\pi imn+ 2\pi i (m \alpha-n\beta))}  e^{\frac{i}{2 \pi} \int J \wedge d\Lambda} Z\;.
\end{align}
The change in the partition function is therefore a pure phase for all $m, n$. However, the transformation \eqref{Lsym} can only account for some of the shift symmetry of the partition function.

The action is not invariant under \eqref{Lsym}, but the change only depends on the sources and theta-characteristics.  We find that, when evaluated on  a winding mode $\phi_{\text{w.m.}}=R(m_\mu+\alpha_\mu)x^\mu/l$, we find that 
\begin{align}
Z\to e^{\frac{2\pi}{r_2}(m(\alpha_0+\alpha_1)- 2r_1r_2n\alpha_1)-2l\sqrt{r_1r_2} n(-J_+^{(0)}+J_-^{(0)})+\frac{2l}{\sqrt{r_1r_2}} m J_-^{(0)}} Z \;,
\end{align}
which agrees with \eqref{phasemain} for the identifications $\alpha = (\alpha_0 + \alpha_1)/r_2$ and $\beta = 2 r_1 \alpha$. This is usually interpreted as saying that the partition function is a section of a line bundle over the space parametrised by ${\cal J}^{(0)} $ modulo the identification (\ref{IntJmain}), that is an auxiliary complex torus parametrised by the zero-modes of the sources, $\mathcal J_{\mathbb T^2}$.

\section{Chiral Two-form Potentials in Six Dimensions}\label{6DPI}

The approach that we used for evaluating the chiral-boson partition function in two dimensions can be extended to $4k+2$-dimensions where one has self-dual $2k+1$ forms. Of particular interest are self-dual three-forms in six dimensions (associated with superconformal field theories such as the (2,0) theory) and self-dual five-forms in ten dimensions (which arise in type IIB string theory). Here, for concreteness,  we will detail the computation in six dimensions, {\it i.e.} $k=1$. The extension to more general $k$ follows readily, and for $k=0$ reproduces the chiral-boson results. 

\subsection{Preliminaries}

Before explicitly computing the partition function let us first make some general comments. Following on from the discussion of Section~\ref{general} we consider the action \eqref{sensources}:
\begin{align}
   S_{\cal C}   = &\frac{1}{(2\pi)^{5}}	\int_{\mathbb{T}^{6}} \left(\frac12 d B\wedge\star_\eta d B -2H\wedge d B +   (H+J)\wedge {  \widetilde{\cal M}}(H+J) + 2H\wedge J- \frac12 J\wedge \star_\eta J \right)\nonumber\\
   & -\frac{1}{(2\pi)^{5}}	\int_{\mathbb{T}^{6}} {\cal C}\wedge dB\ .
\end{align}
We have opted to replace $P$ and $Q$ for the more traditional $B$ and $H$ as in \cite{Lambert:2019diy,Andriolo:2020ykk} and in the second line we have included a boundary term determined by a closed three-form ${\cal C}$, $d{\cal C}=0$. Once again, such a term does not contribute to the equations of motion and preserves all the symmetries of the original theory. In particular, the action is invariant under infinitesimal diffeomorphisms $x^\mu \to x^\mu +\xi^\mu (x)$, provided $\xi^\mu$ is single valued on $\mathbb{T}^{6}$. Analogously to what happened in the previous sections, we will see that the role of ${\cal C}$ is to allow for more general theta-characteristics. 

Next we consider the transformations (\ref{Lsym}).
Under such a transformation we find that $S_{\cal C}$ becomes:
\begin{align}
	S_{\cal C}[B,H, J] \to  S_{\cal C}[B,H,  J] +\frac{1}{(2\pi )^5}\int_{\mathbb T^6} d\Lambda\wedge (dB+{\cal C}-  J)\;. 
\end{align}
These transformations play three roles. First, if $\Lambda=d\lambda$ then we have a familiar abelian gauge transformation $B\to B+ d\lambda$ and the action is invariant. These represent redundancies and, for example,  we can choose $\lambda$ to set  the timelike components of $B$ to zero. Note that in the case of a two-dimensional chiral boson there are no such gauge symmetries. 

Second, if $\Lambda$ is closed ($d\Lambda=0$) but not exact then these transformations are genuine symmetries of the action that we can orbifold by.  This allows us to introduce the analogues of the winding modes from the chiral boson case.  Here $B$ is not single-valued over the torus. Rather we allow for  so-called large gauge transformations
\begin{align}
B\to B + \Lambda\;,\qquad d\Lambda=0	\;,
\end{align}
where  $\Lambda\ne d\lambda$. With this in mind if we go around a loop in the $x^\mu$ direction we can take 
\begin{align}\label{Bwinding}
B(x^\mu+2\pi l  ) = B(x^\mu)+ \Lambda^{(\mu)}\;,
\end{align} 
where $\Lambda^{(\mu)}$ is a closed 2-form. Although this looks like a valid identification for any choice of $\Lambda^{(\mu)}$ one needs to be more careful: We do not want to say that any closed $\Lambda^{(\mu)}$ is allowed, as this condition is too strong.\footnote{In the case of a chiral boson such a condition would amount to saying that $\phi\cong\phi + 2\pi R$ for {\it any} constant $R$.} Rather, we want to impose a flux-quantisation condition
\begin{align}
\frac{1}{(2\pi  )^3}\int_{\Sigma_3} dB	\in R^3\;{\mathbb Z}\;,
\end{align} 
for some fixed $R$ and any three-cycle $\Sigma_3$. This integral will be non-zero if $B$ is not single valued as in (\ref{Bwinding}).
 By including ${\cal C}$ we can be a little more general.  We can  change the flux-quantisation condition to 
\begin{align}\label{Bflux}
\frac{1}{(2\pi   )^3}\int_{\Sigma_3} (dB-{\cal C})	\in R^3\;{\mathbb Z}\;.
\end{align}
We can also interpret these boundary conditions as acting on  the Wilson surface operators of $B$:
\begin{align}
W(\Sigma_2) = e^{\frac{i}{R^3}\frac{1} {(2\pi )^2} \int_{\Sigma_2} B }\ ,    
\end{align}
where $\Sigma_2$ is a 2-cycle. By construction $W(\Sigma_2)$ only depends on the coordinates $x^{\mu'}$ that are normal to $\Sigma_2$. Our boundary condition corresponds to
\begin{align}
W(\Sigma_2)(x^{\mu'} +2\pi l n^{\mu'} ) = e^{2\pi i   \int _{\Sigma_3}{\cal C}}W(\Sigma_2)(x^{\mu'}) \ ,
\end{align}
where $\Sigma_3$ is the 3-cycle obtained by transporting $\Sigma_2$ around the closed loop created by $x^{\mu'}\to x^{\mu'} +2\pi l n^{\mu'}$. 

The third application of the transformation  (\ref{Lsym}) enables us to think of the source zero-modes  as coordinates on an auxiliary complex torus as in Section \ref{general}, $\mathcal J_{\mathbb T^6}$ \cite{Witten:1996hc}. This corresponds to considering transformations where $d\Lambda \ne 0 $, so $J$ shifts. The action is no-longer invariant under \eqref{Lsym} but if we impose another flux-quantisation condition
\begin{align}\label{Jquant}
	\frac{1}{(2\pi)^3 }\int_{\Sigma_3} d\Lambda \in  R^{-3}\;{\mathbb Z}\ ,
\end{align}
then the shift in the action is 
\begin{align}
	S_{\cal C}[B,H,J]&\to  S_{\cal C}[B,H,  J]+ \Delta S_{\cal C}[J]+ 2\pi n\ ,
	\end{align}
	where   $n\in {\mathbb Z}$ and 
	\begin{align}
	\Delta S_{\cal C}[ J]   &= \frac{1}{(2\pi )^5}\int_{\mathbb{T}^{6}} d\Lambda\wedge ( 2{\cal C}-J) \ . 
\end{align}
Thus the path integral and partition function is invariant under such shifts, up to a phase factor that only depends on ${\cal C}$ and $ J$. This is often summarised by saying that the partition function is a section of a line bundle over $\mathcal J_{\mathbb T^6}$, which consists of sources $ J$ modulo $d\Lambda$ subject to (\ref{Jquant}). However, once again one must be a little more careful. Under such a shift we have 
\begin{align}
Z[ J+d\Lambda] & = \int [\mathcal DB'\mathcal DH'] e^{i S_{\cal C}[B',H', J+d\Lambda]}\nonumber\\
& = 	 \int [\mathcal DB'\mathcal DH'] e^{i S_{\cal C}[B+\Lambda,H-(d\Lambda)_+), J+d\Lambda]}\nonumber\\
&=\int [\mathcal D B'\mathcal DH'] e^{i S_{\cal C}[B,H,J]+i\Delta S_{\cal C}[ J]}\nonumber\\
&=e^{i\Delta S_{\cal C}[ J]}\int [\mathcal DB\mathcal DH] e^{i S_{\cal C}[B,H, J]}\nonumber\\
& = e^{i\Delta S_{\cal C}[J]}Z[ J]\ ,
\end{align}
where in the second line we chose $B'=B+\Lambda$, $H'=H-(d\Lambda)_+$ and in the fourth line we assumed $\mathcal [\mathcal DB'\mathcal DH']=[\mathcal DB\mathcal DH]$. When we integrate out $B$ and $H$ we can only perform shifts by   $\Lambda$ if the latter  preserves the form of $B$ and $H$ that we integrated over. Since we integrated over all $H$'s this is not a problem for $H$. However we did not integrate over all $B$'s; rather we restricted to $B$'s that satisfy (\ref{Bflux}). The change $B\to B'$ amounts to just shifting the sum over the winding modes. Thus we can only shift by $\Lambda$'s such that
\begin{align}\label{con1}
\frac{1}{(2\pi   )^3}\int_{\Sigma_3} d\Lambda	\in R^3{\mathbb Z}\ ,
\end{align}
in addition to (\ref{Jquant}). This in turn is only possible if $R^6=r_1/r_2$ is rational, in which case we take
\begin{align}
\frac{1}{(2\pi   )^3}\int_{\Sigma_3} d\Lambda	\in \sqrt{r_1r_2}\ {\mathbb Z}\ .
\end{align}

\subsection{Setup of the Calculation}

Having discussed the six-dimensional action in general let us now calculate the partition function. Here we can be more explicit with our discussion.
To that end we introduce  a basis of (anti)self-dual 3-forms as in \eqref{selfdualbasis}
\begin{align} 
\omega^A_+ &= (1+\star_\eta)dx^0\wedge dx^i\wedge dx^j\nonumber	\\
\omega_{-A} & = (1-\star_\eta)dx^0\wedge dx^i\wedge dx^j\ ,
\end{align}
where $i,j=1,\ldots,5$ and $A=(i j)$ with $i<j$ running over all $10$ possibilities. Note that these are chosen such that
\begin{align}
\omega_+^A\wedge \omega_{-B}	 = 2\delta^A_B d^6x\;.
\end{align}
Furthermore we introduce a basis of 2-forms:
\begin{equation}
\omega_2^a  = \left\{	\begin{array}{ll}
	dx^0\wedge dx^i \\ dx^i\wedge dx^j
\end{array} \right.\;,
\end{equation}
where the index $a$ runs over $(0i)$ and $(i j)$ (again with $i<j$). Thus there are $5+10=15$ values of $a$. It is helpful to expand
\begin{align} dx^\mu \wedge \omega_2^a= K^{\mu a}{}_B \omega^B_+ + L^{\mu a B}{} \omega_{-B} \;,
\end{align}
for some $K^{\mu a}{}_B $ and  $L^{\mu a B}{}$.
In particular we see that
\begin{align}
K^{\mu a}{}_A & = \frac{1}{2(2\pi l)^6}\int_{\mathbb{T}^{6}}  dx^\mu \wedge \omega^a_2\wedge \omega_{-A}	\nonumber\\
 L^{\mu a A} & = -\frac{1}{2(2\pi l)^6}\int_{\mathbb{T}^{6}}  dx^\mu \wedge \omega^a_2\wedge \omega_+^{A}\ .
 \end{align}
For future reference we observe that the non-vanishing values of $K^{\mu a }{}_B,L^{\mu a B}$ are $\pm \frac12$. However the non-vanishing components of $K^{\mu  a B}\pm L^{\mu a}{}_B$ are $\pm1$.
 Furthermore we can compute
\begin{align}
( dx^\mu \wedge \omega^a_2)\wedge \star_\eta(	dx^\nu \wedge \omega^b_2) &=-2( K^{\mu a}{}_BL^{\nu b B}{} +K^{\nu b}{}_BL^{\mu a B}{})d^6x\nonumber\\
&=: 2\kappa^{\mu\nu ab} d^6x\ .
\end{align}

Next, we need to construct the matrix $\widetilde {\cal M}^{AB}$ as in \eqref{Mform}. This is rather cumbersome for a general metric. However, we can make the following important observation. To integrate out $H_A$ from the path integral we need to ensure that $\mathrm{Im}(\widetilde {\cal M}^{AB})>0$. We now prove that this is the case if one chooses the branch $\sqrt{-1}=-i$, as for the 2D chiral boson. With this choice the Lorentzian Hodge operator $\star_{g_{\epsilon= 2i}}$ for the Euclidean metric $g_{\epsilon=2i}$ is related to the Euclidean Hodge operator $\star^E_{g_{\epsilon= 2i}}$ through $\star_{g_{\epsilon= 2i}}=-i\star^E_{g_{\epsilon= 2i}}$. Recall that 
\begin{align}
\langle \Omega \mid \Omega'\rangle := \int_{\mathbb{T}^{6}} \Omega^*\wedge \star
^E\Omega'\ ,
\end{align}
defines a positive-definite inner-product over the space of (possibly complex-valued) three-forms in 6D, for any Euclidean metric. Consider now a non-vanishing three-form $\Omega = \Omega_A\varphi^A_+$, self-dual with respect to $\star_{g_{\epsilon= 2i}}$. Then
\begin{align}
0<\langle \Omega \mid \Omega\rangle 	&=\int_{\mathbb{T}^{6}} \Omega^*\wedge \star^E_{g_{\epsilon=2i}}\Omega\nonumber\\
&=i\int_{\mathbb{T}^{6}} \Omega^*\wedge \star_{g_{\epsilon=2i}}\Omega\nonumber\\
&=i\int_{\mathbb{T}^{6}} \Omega^*\wedge \Omega\nonumber\\
	&=i(2\pi l)^6\Omega^*_A\Omega_B({\cal N}^*{\cal K}^T - {\cal K}^*{\cal N}^T)^{AB}\nonumber\\ &= -2i(2\pi l)^6\Omega^*_A\Omega_B({\cal N}^*\widetilde{\cal M}^T{\cal N}^T-{\cal N}^*\widetilde{\cal M}^*{\cal N}^T )^{AB}\nonumber\\
&=-2i(2\pi l)^6\Omega_A\Omega^*_B({\cal N}(\widetilde{\cal M}-\widetilde{\cal M}^\dag){\cal N}^\dag )^{AB}\nonumber\\
	&= -2i (2\pi l)^6({\cal N}^\dag{\Omega}^*)^\dag(\widetilde{\cal M}-\widetilde{\cal M}^\dag)({\cal N}^\dag\Omega^*)\ ,
\end{align}
where we remind the reader that 
$\varphi^A_+ =     {\cal N}^A{}_B\omega^B_+ + {\cal K}^{AB}\omega_{-B}$ and  ${\cal K}^{AB} = -{\cal N}^{A}{}_C \widetilde{\cal M}^{CB}$.
Thus we find that choosing $\sqrt{-1}=-i$ leads to $-i(\widetilde{\cal M}-\widetilde{\cal M}^\dag)^{AB}$ being positive definite. This implies that
\begin{align}
{\rm Re}\left( i H_AH_B\widetilde{\cal M}^{AB}\right)  &< 0\ ,
\end{align}
for any real values of $H_A$, which will be needed to ensure convergence of the functional integrals appearing in the partition function.

To continue we expand the fields, sources and ${\cal C}$ as
\begin{align}
H & = H_A\omega^A_+\nonumber\\
B &  = B_a\omega^a_2\nonumber\\
 J & = J^+_{A}\omega^A_+ + J^{-A}\omega_{-A}\nonumber\\
{\cal C} &  = \alpha_{a\mu  } dx^\mu \wedge \omega^a_2\ ,	
\end{align}
where $H_A$ and  $B_a$ are functions and $\alpha_{\mu a}$ constants. The action can now be written as
\begin{align}\label{action6D}
S_{\cal C}&= \frac{2}{(2\pi)^5}\int_{{\mathbb T}^6} d^6x\Big( \frac12 \kappa^{\mu\nu ab}\partial_\mu B_a\partial_\nu B_b -2L^{\mu a A}  H_A \partial_\mu B_a+  (H_A+J^+_{A})(H_B+J^+_{B})\widetilde{\cal M}^{AB}\nonumber\\ 
&\qquad\qquad\qquad\qquad +2H_A J^{-A}  +J^+_{A}J^{-A}  -(K^{\mu a A}L^{\nu b}{}_A-L^{\mu a}{}_AK^{\nu b A})\alpha_{a\mu }\partial_\nu B_b\Big)\nonumber\\
& = \frac{2}{(2\pi)^5} \int_{{\mathbb T}^6} d^6x \Big(\frac12 G^{\mu\nu ab}\partial_\mu B_a\partial_\nu B_b +2L^{\mu a A}\partial_\mu B_a(J^+_{A}+\widetilde{\cal M}^{-1}_{AB}J^{-B})+  \hat H_A\hat H_B\widetilde{\cal M}^{AB}\nonumber\\
	&\qquad\qquad\qquad\qquad -(K^{\mu a A}L^{\nu b}{}_A-L^{\mu a}{}_AK^{\nu b A})\alpha_{a\mu  }\partial_\nu B_b -J^+_{A}J^{-A} -J^{-A}J^{-B}\widetilde{\cal M}^{-1}_{AB}
 \Big)\ ,
\end{align}
where
\begin{align}
\hat H_A = H_A + J^+_{A} - (L^{\mu a B}\partial_\mu B_a-J^{-B})\widetilde{\cal M}^{-1}_{AB}\ ,
\end{align}
and\begin{align}
G^{\mu\nu ab} &= \kappa^{\mu\nu ab} - L^{\mu a C}L^{\nu b D}\widetilde{\cal M}^{-1}_{CD}	\nonumber\\
&=-K^{\mu a}{}_BL^{\nu b B}{}-K^{\nu b}{}_BL^{\mu a B}{}  - 2L^{\mu a C}L^{\nu b D}\widetilde{\cal M}^{-1}_{CD}\ .
\end{align} 

\subsection{Explicit Evaluation}

At this stage we are ready to evaluate the path integral for \eqref{action6D}. In principle, one would have to carry this out via the Dirac path-integral procedure, as we implemented for the chiral boson in Section~\ref{sec:dirac}. However, encouraged by the observation in two dimensions that the Dirac and the Feynman path-integral prescriptions led to the same answer (up to multiplicative constants), we will be cavalier and proceed directly with the Feynman path integral
\begin{align}
    Z[J^+_{A},J^{-A}, \widetilde {\mathcal M}^{AB}] = \int[\mathcal DB \mathcal D \hat H ]e^{i S_{\mathcal C}}\;.
\end{align}
It should be noted that one should also mod-out by the volume of the gauge symmetry of the  2-form $B$; for the sake of simplicity we will work in the gauge where $B_{0i}=0$, so there is no infinite-dimensional residual gauge and we do not need to introduce ghosts fields. 

We can perform the $\hat H_A$ functional integrals since we have seen that $-i(\widetilde {\cal M}-\widetilde {\cal M}^\dag)^{AB}$ is positive definite so that they are convergent. There are ten values of $A$ and each $\hat H_A$ has an  expansion in terms of a real zero-mode and an infinite tower of complex Fourier modes: 
\begin{align}
\hat H_A & = h_{0A} +\sum'_{n_{\mu}} h_{n_\mu A}e^{in_{\mu} x^\mu /l}	 \;,
\end{align} 
with $(h_{A,n_\mu})^*=  h_{A,-n_\mu}$. However, these can be seen to integrate to one after zeta-function regularisation, as we saw above in (\ref{regularised}).

Thus we are left with evaluating
\begin{align}\label{wJ}
Z[J^+_{A},J^{-A}, \widetilde {\mathcal M}^{AB}]  \sim \int [\mathcal DB] e^{iS_{eff}}\ ,
\end{align}
where the effective action left over as the result of the $\hat H_A$ integration is given by
\begin{align}
	S_{eff}  &= \frac{2}{(2\pi)^5} \int_{{\mathbb T}^6} d^6x \Big(\frac12 G^{\mu\nu ab}\partial_\mu B_a\partial_\nu B_b +2L^{\mu a A}\partial_\mu B_a{\cal J}_{A}-(K^{\mu a A}L^{\nu b}{}_A-L^{\mu a}{}_AK^{\nu b A})\alpha_{\mu a}\partial_\nu B_b\nonumber \\ &\qquad\qquad\qquad\qquad\qquad-{\cal J}_{A} \left(\widetilde{\cal M}^*(\widetilde{\cal M}^*-\widetilde{\cal M})^{-1}\widetilde{\cal M}\right)^{AB}({\cal J}_B- {\bar {\cal J}}_B)	\Big)\;,    
\end{align}
and  we have introduced
\begin{align}
{\cal J}_{A} &= J^+_{A} + \widetilde{\cal M}^{-1}_{AB}J^{-B}\nonumber\\
\bar {\cal J}_A &=J^+_{A} + J^{-B} (\widetilde {\cal M}^{-1})^\dag_{BA}\ .\end{align}
 This defines a complex structure on the sources where we view ${\cal J}_{A}$ and $\bar {\cal J}_A$ as  holomorphic and anti-holomorphic coordinates respectively.

It is clear from the simple dependence of $S_{eff}$ on $\bar {\cal J}_{A}$ that the partition function is holomorphic in a twisted sense:
\begin{align}
{\bar {\cal D}}^{A}Z[{\cal J}_{A},\bar {\cal J}_B]\:=\left(\frac{\delta}{\delta  \bar {\cal J}_A}-2\pi i l^6\left(\widetilde{\cal M}^*(\widetilde{\cal M}^*-\widetilde{\cal M})^{-1}\widetilde{\cal M}\right)^{AB}{\cal J}_{B} \right)Z[{\cal J}_{A},\bar {\cal J}_B]=0	\;,
\end{align}
while the holomorphic derivative is 
\begin{align}
{\cal D}^B = 	\frac{\delta}{\delta {\cal J}_{B}}+2\pi i l^6\left(\widetilde{\cal M}(\widetilde{\cal M}-\widetilde{\cal M}^*)^{-1}\widetilde{\cal M}^*\right)_{AB}{\bar {\cal J}}_{B}\;.
\end{align}
 This holomorphic structure  captures the simple dependence on $\bar {\cal J}_B$:
\begin{align}
Z[{\cal J}_{A},\bar{\cal J}_B] = e^{\frac{2i}{(2\pi)^5} \int_{{\mathbb T}^6} d^6x  \left(\widetilde{\cal M}^*(\widetilde{\cal M}^*-\widetilde{\cal M})^{-1}\widetilde{\cal M}\right)^{AB}{\cal J_A}{\bar {\cal J}}_B}	  Z_h[{\cal J}_{A}]	\ ,
\end{align}
where $Z_h[{\cal J}_A]$ can also be further factorised as
\begin{align}
Z_h[{\cal J}_{A}]	= &e^{-\frac{2i}{(2\pi)^5} \int_{{\mathbb T}^6} d^6x  \left({\cal \widetilde M}^*(\widetilde{\cal M}^*-\widetilde{\cal M})^{-1}\widetilde{\cal M}\right)^{AB}{\cal J_A}{{\cal J}}_B}Z_h^{(0) }[{\cal J}_{A}]\ .
\end{align}
So our final task is to compute
\begin{align}
    Z_h^{(0)}[{\cal J}_{A}] &= \int [\mathcal DB]e^{i S^{(0)}_{eff}[B,{\cal J}_A]}\nonumber\\
    S^{(0)}_{eff}& = \frac{2}{(2\pi)^5} \int_{{\mathbb T}^6} d^6x \Big(\frac12 G^{\mu\nu ab}\partial_\mu B_a\partial_\nu B_b +2L^{\mu a A}\partial_\mu B_a{\cal J}_{A}\nonumber\\ &\qquad\qquad\qquad\qquad-(K^{\mu a A}L^{\nu b}{}_A-L^{\mu a}{}_AK^{\nu b A})\alpha_{\mu a}\partial_\nu B_b 	\Big)\ .
\end{align}
To this end we expand the fields in Fourier modes
\begin{align}
B_a &= b_a + {w_{a\mu}}{} x^\mu +\sum'_{n_{a\mu}} b_{a,n_{a\mu}}e^{in_{a\mu} x^\mu /l}\qquad	n_{a\mu} \in {\mathbb Z}\ ,
\end{align}
with  $(b_{a,n_{a\mu}})^*= b_{a,-n_{a\mu}}$. We have separated out the zero-modes as they are real. The $w_{a \mu}$ are the analogues of winding modes and the flux-quantisation condition (\ref{Bflux}) implies that
\begin{align}\label{wis}
w_{a\mu} = \frac{R^3}{l^3}\left( m_{a\mu} + \alpha_{a\mu }    \right)\qquad m_{a\mu} \in {\mathbb Z} \ .
\end{align}
Since the action is quadratic the evaluation of the partition function factorises into a contribution arising  from a sum over the winding modes $m_{a\mu}$ and an integral over the oscillator modes $b_{a n_\mu}$:
\begin{align}
    Z_h^{(0)}[{\cal J}_{A}] = Z^{(0)}_{\text{w.m.}} Z^{(0)}_{\rm osc}\ .
\end{align}

 Let us first evaluate the action on the oscillator modes. 
The calculation is similar in form to that of Section~\ref{sec:osc}. The action evaluates to
\begin{align}
S_{\text{eff} }^{(0)}= -2\pi l^4&\sum_{a, n_{\mu }}\Big(G^{\mu\nu ab}n_\mu n_\nu b_{a n_{\mu }}  b_{b,-n_{\nu}}   - {4i}{l} L^{\mu a A}n_{a\mu }b_{an_{ \mu }}{\cal J}_{+A}^{-(n_{ \mu})}\Big)	\ ,
\end{align} 
where the integral over ${\mathbb T}^6$ has imposed $n_{a\mu}=-n_{b\mu}=n_{\mu}$. The integral over the $b_{an_\mu}$'s now produces
\begin{align}
    Z_{\rm osc}&\sim  \prod'_{n_\mu}\frac{-i l^{-4}}{\det (G^{\mu\nu ab}n_\mu n_\nu)}\nonumber\\
    &\sim \prod'_{n_\mu}\frac{1}{ \det (2K^{\mu a}{}_BL^{\nu b B}{}n_\mu n_\nu+2L^{\mu a A}L^{\nu b B}\widetilde{\cal M}^{-1}_{AB}n_\mu n_\nu)}\nonumber\\
    &\sim \prod'_{n_\mu}\frac{1}{\det (2L^{\nu b A}{}(L^{\mu a A} -K^{\mu a}{}_A)n_\mu n_\nu+4L^{\mu a A}L^{\nu b B}T_{AB}n_\mu n_\nu)}
    \ ,
\end{align}
where the determinant is over the $a,b$ indices and \begin{align}
    { T}_{AB}= -\frac12({\delta}_{AB}+\widetilde{\cal M}^{-1}_{AB})\ .
\end{align}  
We should also be careful here to impose a gauge-fixing condition such as only including modes where $a=A$. It is difficult to evaluate this expression more precisely in general. We recall that the non-zero values of ($L^{\mu a A}-K^{\mu a}{}_A)$ and $2L^{\mu a A}$ are $\pm1$ so the determinant is of a matrix which is quadratic in the integers $n_\mu$ and linear in $T_{AB}$, much like  (\ref{zetareg}). We will suggestively denote it as:
\begin{align}
  Z_{\rm osc}
    :=  \frac{1}{\eta^{10}_{6D}({  T}_{AB})}\ .
\end{align}
 
Next, we  evaluate the action on the winding modes. Again it is helpful to introduce
\begin{align}
w^+_A  
& := K^{\mu a}{}_Aw_{\mu a}\nonumber\\
w^{-A} 
& := L^{\mu a A}w_{\mu a}	\ .
\end{align}
We can apply similar maps to $\alpha_{a\mu  }$ and $m_{a \mu  }$. 
In this case we find
\begin{align}
S^{(0)}_{\text{eff} }=  \frac{2}{(2\pi)^5} \int_{{\mathbb T}^6} d^6x &\Big(  -w^{-A}w^+_A -   w^{-A}w^{-B}\widetilde{\cal M}^{-1}_{AB}    -\frac{R^3}{l^3}\alpha^{+}_{A}w^{-A} +\frac{R^3}{l^3}\alpha^{-A}w^+_A+2w^{-A}{\cal J}_{A} \Big)\nonumber\\
= \frac{2}{(2\pi)^5} \int_{{\mathbb T}^6} d^6x &\Big(  -  w^{-A}w^{-B}\widetilde{\cal M}^{-1}_{AB}  +2w^{-A}({\cal J}_{A}-\frac12(R/l)^3\alpha^+_A)   -(w^{-A}- (R/l)^3\alpha^-_A)w^+_A \Big)\nonumber\\
= \frac{2}{(2\pi)^5} \int_{{\mathbb T}^6} d^6x &\Big(- w^{-A} w^{-B}(\delta_{AB}+\widetilde{\cal M}^{-1}_{AB})  +2 w^{-A}\left({\cal J}_{A}+(R/l)^3( \alpha^{-A}-\alpha^{+}_{A})\right) \nonumber\\
& +( w^{-A}- (R/l)^3 \alpha^{-A}) w'_A    +(R/l)^6\alpha^{-A}(\alpha^+_A-\alpha^{-A})\Big)\;,
\end{align}
where we have introduced
\begin{align}
 w'_A &= w^{-A}-w^+_A +\frac{R^3}{l^3}(\alpha^+_A-\alpha^{-A})\ .
\end{align} 
The point about   $w'_A$ is that, given (\ref{wis}),   then 
\begin{align}
w'_A = w^{-A}-w^+_{A} =  (R/l)^3m_A\ ,
\end{align}
for some $m_A\in  {\mathbb Z} $.
Whereas
\begin{align}
 w^{-A}- (R/l)^3 \alpha^{-A} =  (R/l)^3m^{-A}\qquad m^{-A}\in  {\mathbb Z}\ .
\end{align}
Therefore we see that the sum over $m_A$ imposes a delta-function constraint
\begin{align}
\sum_{m_A\in {\mathbb Z}^{10}} e^{4\pi i R^6  m^{-A}m_A}	=   \sum_{p^A\in {\mathbb Z}^{10}}\delta(2R^6 m^{-A}-p^A)\ .
\end{align}
For $R^6$ irrational, the only solution is $ m^{-A}=p^A=0$
and hence $Z^{(0)}_{\text{w.m.}}\sim 1$. However, for $R^6=r_1/r_2$ we find  
$m^{-A} = r_2 n^A/2$ and $p^A=r_1n^A$ (recall $m^{-A}\in \frac 12 \mathbb Z$). Substituting this back into the action gives 
\begin{align}
Z^{(0)}_{\text{w.m.}} & =  e^{4 \pi i R^6 \alpha^{-A}(\alpha^+_A-\alpha^{-A})}\sum_{p^A}\sum_{m^{-A}}\delta(2R^6m^{-A}-p^A)\nonumber\\
& 
\times e^{-4\pi \pi R^6 i (m^{-A}+\alpha^{-A})(m^{-B}+\alpha^{-B})(\delta_{AB}+\widetilde{\cal M}^{-1}_{AB}) + 8\pi R^3 i (m^{-A}+\alpha^{-A})( {\cal J}_A^{(0)}+R^3( \alpha^{-A}-\alpha^{+}_{A}))}
\nonumber\\
& \sim    e^{-\pi i \alpha^A\beta_A}\Theta \left[\begin{array}{c}
\alpha^A \\ \beta_A 	
\end{array}\right]
\left(\sqrt{r_1r_2} {\cal J}^{(0)}_{A}\mid 2r_1r_2{ T}_{AB} \right)
\;,
\end{align}
where we find the higher-dimensional theta functions:
\begin{align}
  &\Theta \left[\begin{array}{c}
\alpha^A \\ \beta_A 	
\end{array}\right]
\left( z_{A}\mid \tau_{AB} \right)   := \sum_{n^{A}\in {\mathbb Z}^{10}}		e^{\pi i  (n^{A}+\alpha^A )(n^{B}+ \alpha^B )\tau_{AB} + 2\pi  i (n^{A}+\alpha^A)(z_A +\beta_A)}\ .
\end{align}
Here we have introduced the normalised source zero-mode
\begin{align} \mathcal{J}^{(0)}_A = \frac{2l^3}{(2\pi l)^6} \int d^6 x \left(J^{+}_A + \widetilde{\cal M}^{-1}_{AB}J^{-B} \right)\ ,
\end{align}
and theta-characteristics
\begin{align}
\alpha^A &= 2\alpha^{-A}/r_2 \nonumber\\
\beta_A &=	2r_1(\alpha^{-A}-\alpha^+_A)  \ .
\end{align} 
In summary, our final answer for the six-dimensional partition function is
\begin{align}
    Z \sim &e^{-\pi i \alpha^A\beta_A}e^{-\frac{2i}{(2\pi)^5} \int_{{\mathbb T}^6} d^6x  \left(\widetilde{\cal M}^*(\widetilde{\cal M}^*-\widetilde{\cal M})^{-1}\widetilde{\cal M}\right)^{AB}{\cal J_A}({\cal J}_B-{\bar {\cal J}}_B)} \frac{\Theta \left[\begin{array}{c}
\alpha^A \\ \beta_A 	
\end{array}\right]
\left( \sqrt{r_1r_2}{\cal J}^{(0)}_{A}\mid 2r_1r_2{  T}_{AB} \right)}{\eta^{10}_{6D}({\cal T}_{AB})}\ .
\end{align}
Note that for vanishing characteristics we find a higher-dimensional analogue of $T$-duality: $R\leftrightarrow 1/R$.
 
It is important to make some comments about gauge symmetries. The expressions for $Z_{\text{w.m.}}$ that we have above are gauge invariant and hence the path integral will over-count the physical degrees of freedom. We have assumed a simple gauge-fixing condition of the form $B_{0i}=0$.
This means that only components of $B = B_a\omega^a_2$ with $a= (i j)$ need to be considered and these are in one-to-one correspondence with the $A$-indices. We do not expect that this leads to a change in our final result as the over-counting simply leads to an overall constant multiplying the partition function. 

One can also see this at the level of the winding modes $w_{a\mu}$. There are $6\times 15=90$ possible choices of $w_{a\mu }$ but only $20$ possible 3-forms $dB$. Indeed one sees that some $w_{a\mu}$ lead to $dB=0$: for each choice of $a$ there are in fact only 4 choices of $\mu$  such that $w_{a\mu} dx^\mu \omega^a$ is non-zero.   Furthermore, different choices of $w_{a\mu }$ can lead to the same non-zero $dB$. Again this represents an over-counting but one which simply leads to an overall rescaling of the partition function. The gauge-invariant information is only contained in the 3-forms and therefore we assume that the sum over $w_{a\mu}$ and integral over oscillating modes $b_{a n_\mu}$ can be entirely captured by the contributions from $(w^+_A, w^{-A})$ and $n^+_A,n^{-A}$. 

To demonstrate this more explicitly, we consider the case where the physical metric is (in  Lorentzian signature)
\begin{equation}
g_{\mu\nu} = \text{diag}\left(-L_0^2, L^2_1, \ldots,L_5^2\right)\;.
\end{equation}
To construct $\widetilde{\cal M}^{AB}$ we see that a basis of self-dual 3-forms with respect to $\star_g$ are given by
\begin{align}
\varphi^A_+ &= (1+\star_g) dx^0\wedge dx^i\wedge dx^j \cr
& = \frac{\omega^A_{+}+\omega_{A-}}{2}
+V L_i^{-2}L_j^{-2}L_0^{-2}\frac{\omega^A_{+}-\omega_{A-}}{2}\cr
& = \frac{1+V/L^2_i L^2_j L_0^2}{2}\omega^A_{+} + 
\frac{1-V/L^2_i L^2_j L_0^2}{2}\omega_{A-}\ ,
\end{align}
where $V = L_0L_1\ldots L_5$. From here we can read off
\begin{align}
\widetilde {\cal M}^{AB} = - 	\frac{1-V/A_AL_0^2}{1+V/A_AL_0^2}\delta^{AB}\ ,
\end{align} 
where $A_A = L_i L_j$. Finally we want to Wick rotate in the same way as in Section \ref{2DPI}, which effectively amounts to sending
\begin{align}
L_0\to -i L_0	\;.
\end{align}
Thus we find
\begin{align}
\widetilde {\cal M}^{AB} &= - 	\frac{1-i V/A_AL_0^2}{1+i V/A_AL_0^2}\delta^{AB}\nonumber\\
&=-\frac{\tau_A+1}{\tau_A-1}\delta^{AB}\ ,
\end{align}
where $\tau_A=i L_0A_A/V$. For such metrics $\widetilde{\cal M}^{AB}$ becomes diagonal and each component has the form that we saw for the chiral boson but with a purely imaginary complex structure $\tau_A$. Note that in this case
 \begin{align}
     {\cal T}_{AB} &=-\frac12\left(\delta_{AB} +\widetilde {\cal M}^{-1}_{AB}\right)\nonumber\\
     &=- \frac{1}{1+\tau_A}\delta_{AB}\ .
 \end{align}
 As a result $Z^{(0)}_{\rm w.m.}$  factorises into a product  of more familiar functions from the two-dimensional chiral boson discussion:
 \begin{align}
  Z^{(0)}_{\rm w.m.} & = \prod_{A=1}^{10} \theta \left[\begin{array}{c}
\alpha^A \\ \beta_A 	
\end{array}\right]\left(\sqrt{r_1r_2}{\cal J}^{(0)}_A\mid -2r_1r_2/(1+\tau_A)\right)
 \end{align}
Lastly, we note that $-1/(1+\tau_A)$ is simply an $S$ and $T$ modular transformation away from $\tau_A$.

\section{Conclusions}\label{conclusions}

In this paper we performed a direct calculation of the partition function associated with the Sen action for chiral forms in $4k+2$ dimensions in a path-integral formulation. As this action contains unphysical fields with the wrong-sign kinetic term, convergence of the path integral was achieved through a non-standard analytic continuation to Euclidean signature {\it via} a complex deformation of the metric and not time. This procedure had the additional benefit of leaving the self-duality condition of the self-dual form, $Q=\star_\eta Q$, untouched and directly led to a holomorphic result.

To appreciate this last point, one should take a step back to understand what happens within the holomorphic-factorisation approach to the partition function of the chiral boson. In that framework, one  starts with the Lorentzian path integral of a non-chiral boson $\phi$, {\it i.e.} $Z_{\rm n.c.}\sim \int [\mathcal D\phi]\exp(i \int d^2x \partial_{\mu}\phi \partial^{\mu}\phi)$ and by Wick rotating as usual, $x^0\to -ix^0$, one  evaluates $\int [\mathcal D\phi]\exp(- \int d^2x(\partial_0 \phi)^2+(\partial_1 \phi)^2)$. This leads to a real result of the schematic form
\begin{align}
  Z_{\rm n.c.} \sim \mathcal W( \tau-\bar \tau )\sum \frac{\theta}{\eta} \frac{\bar{\theta}}{\bar \eta}\ ,
\end{align}
where the sum is over the characteristics of the theta functions, see \cite{Henningson:1999dm}, and $\tau $ is the complex structure of   the torus. Then, one would like to conclude that the chiral-boson partition function is indeed the holomorphic  theta function, with some undetermined characteristics (which can get fixed, case by case, according to the actual physical system that the chiral boson is meant to describe). In so doing one also needs to ignore the anomalous factor $\mathcal W$.

Instead, in the approach taken here, the kinetic term of the non-chiral boson does not get Wick rotated and one computes the path integral of a \textit{complex} action. Rather, the convergence arises from the $ \widetilde {\cal M} $ term in the action.  
In fact,  thanks to implementing the Wick rotation as a metric deformation, and to the precise nature of the non-standard coupling of the Sen action to the curved background, $\widetilde {\cal M}$ is simply related to the torus complex structure  $\tau$  and enters in a manifestly holomorphic way. 

In this way, for  the chiral boson in two dimensions, we  reproduced the classic $\theta/\eta$ result by a  calculation of the path integral on $\mathbb T^2$. The argument of this expression was an $SL(2,\mathbb Z)$ transformation away from the usual $\mathbb T^2$ complex structure. General theta-function characteristics were incorporated by introducing a topological term to the Sen action and adjusting the periodicities of the scalar on the torus. We then proceeded to repeat the same computation for the significantly more complicated case of the six-dimensional theory on $\mathbb T^6$, under certain assumptions about the equivalence of the Dirac and Feynman path-integral prescriptions for the Sen action and the form of the contributions of 2-form gauge orbits. The result, which was a generalisation of the two-dimensional one, can be straightforwardly extended to higher $k$. 

It is worth making contact between our calculation and the canonical-quantisation computation of the partition function of the Sen action. In the latter approach one introduces a pair of non-canonically conjugate variables, $\Pi^{\pm}\sim \Pi^{P}\pm dP$  (see \cite{Andriolo:2020ykk,Sen:2019qit}), in terms of which the Hamiltonian  schematically splits into $\mathcal{H}\sim \mathcal{H}_++\mathcal{H}_-$. Here $\mathcal{H}_+$ is a negative-definite Hamiltonian which completely decouples from the system, while $\mathcal{H}_-$ is the physical Hamiltonian which describes the correct spectrum of the chiral form. Therefore, within the canonical approach to quantisation, one can simply recover the partition function of, {\it e.g.}~, the compact chiral boson by computing $\text{Tr}e^{2\pi i \tau R \mathcal{H}_-}$; see \cite{Sen:2019qit}. Note that, due to the nature of the Legendre transform, the decoupling of the unphysical modes is not straightforward in the Lagrangian formulation of the theory. Nevertheless we  found a sensible result by keeping all modes and this is due to the non-standard Wick rotation, which preserved the wrong-sign modes in Lorentzian signature as oscillatory contributions in the Euclidean path-integral. In other words,  the Wick rotation \eqref{MwickRotated} left the kinetic term of $\phi$ in  \eqref{eq:Sfull} unaffected and thus it made sense to compute the path-integral of the Wick-rotated theory without removing any contributions.

It is satisfying to see the chiral partition function emerge directly from an honest functional-integral calculation. Our work hence provides nontrivial evidence that the proposal \cite{Sen:2015nph,Sen:2019qit} correctly captures the physics of chiral forms also within the framework of the path-integral approach to quantisation. It would be interesting to apply the above methods to more complicated background geometries, such as higher-genus Riemann surfaces in two dimensions for which a globally defined timelike killing vector does not exist.

Finally, we note that although the action is invariant under infinitesimal diffeomorphisms the partition function is not modular invariant, indicating that there is a failure of global diffeomorphisms. A failure of modular invariance implies that the chiral field in $4k+2$ dimensions is not a genuine $4k+2$ dimensional system; it is more appropriate to think of it in terms of the dynamics on the boundary of a $4k+3$ Chern--Simons theory. However, since we reproduced the correct result from a  path-integral formulation in $4k+2$ dimensions, perhaps it could be possible to relate the Sen action to Chern--Simons theory in one dimension higher. Another hint towards this direction is the fact that the fields $Q$ and $P$ mix under diffeomorphisms, \eqref{DiffeoTranf}, which could be a sign that they both originate from the same object in a higher-dimensional theory. A higher-dimensional interpretation of the Sen action could additionally shed some light on the true geometric nature of $\widetilde{\cal M}$. We hope to return to these questions in the future.

\section*{Acknowledgements}

We would like to thank S.~Andriolo, D.~Berman, M.~Bullimore, A.~Grigoletto, E.~Harris, S.~Murthy, S.-H.~Shao, G.~Watts and especially A.~Sen for useful discussions and comments. E.A. is supported by the Royal Society RGF\textbackslash EA\textbackslash 180073 and would like to thank the Simons Center for Geometry and Physics for hospitality while this work was being completed.  T.O. is supported by the STFC studentship ST/S505468/1. C.P. is partially supported by the Royal Society URF\textbackslash R\textbackslash 180009 and the STFC  ST/P000754/1.

\begin{appendix}

\section{Detailed evaluation of 2D Partition Function}\label{app:details}

In this appendix we present the calculation of the 2D partition function of Sections~\ref{sec:osc} and \ref{general} in full detail. These algebraic manipulations are very familiar from Conformal Field Theory (see e.g. \cite{DiFrancesco:1997nk}) but their applications and interpretation in the context of the Sen action are somewhat different. Our modular function conventions are:
\begin{align}\label{thetas}
	\theta \! \begin{bmatrix} \vspace{-20pt} \\ \alpha  \\ \vspace{-20pt} \\ \beta \\ \vspace{-20pt} \end{bmatrix} \! ( z | \tau ) := \sum_{p\in{\mathbb Z}}
	e^{i\pi (p+\alpha)^2 \tau +2i\pi (p+\alpha)(z+\beta)}\;,\quad \mathrm{Im}(\tau)>0\;, \, z \in \mathbb{C}\;,  \, \alpha, \beta \in \mathbb{R}\;,  
	\end{align}
and
\begin{align}
\eta(T) = e^{\pi i T/12}\prod_{n=1}^\infty (1- e^{2\pi in T }) \ .
\end{align}

\subsection{Oscillator Modes}\label{app:osc}

Evaluating the effective action \eqref{Sen'sEffective} on the oscillator modes we find
\begin{align}
S_{\text{eff}}[\phi_{\text{osc}},J^\pm]&=  \pi \sum_{n_0,n_1}^\prime  \left\{ |\phi_{n_0,n_1}|^2 \left( n_1^2-n_0^2 -  \frac{1}{{\mathcal M}}(n_0+n_1)^2\right)  \right.\nonumber\\
&\left.\qquad\qquad\qquad
+4i l(n_0+n_1) \phi_{n_0,n_1}\big( J^+_{-n_0,-n_1} + \frac{1}{{\mathcal M}} J^-_{-n_0,-n_1} \big) \right\}\ .
\label{oscillationS}
\end{align}
Even though the source $J$ explicitly enters this expression, the dependence on $J$ will be washed away upon integrating over each oscillator, and the contribution $Z_{\text{osc}}$ of the oscillatory modes to the partition function will be independent of $J$. Indeed, let $\phi=\phi_1+i\phi_2$ represent a generic oscillator mode appearing in \eqref{oscillationS} and let ${j}$ denote a contribution from the  source. For a complex number $G$ with positive imaginary part we then have
 \begin{align}
 \int_{\mathbb{C}} d\phi d \phi^* e^{i G|\phi|^2+i{j}\phi} & = 2 \int_{\mathbb{R}^2}  d\phi_1d \phi_2 e^{i G(\phi_1^2+\phi_2^2)+i{j}(\phi_1+i\phi_2)}\nonumber\\
  	& =  2\int_{\mathbb{R}}  d \phi_1 e^{i G\left(\phi_1+\frac{ j}{2G}\right)^2}\int_{\mathbb{R}}  d \phi_2 e^{i G\left(\phi_2+i\frac{ j}{2G}\right)^2}\nonumber\\
  	& =\frac{2\pi i}{G}\;.
 \end{align}
 Note that the chiral (right-moving) modes with $n_0+n_1=0$ lead to a vanishing action: since each of them merely contributes to the path integral as an overall infinite constant ($\sim \int d\phi d \phi^* e^{i0}$), we will only include  the modes $\phi_{n_0,n_1}$ with $n_0+n_1\ne 0$. Furthermore, due to the reality condition $(\phi_{n_0,n_1})^* = \phi_{-n_0,-n_1}$, we restrict to $n_0+n_1> 0$ to avoid double counting. Therefore, $Z_{\text{osc}}$ evaluates to 
\begin{align}
Z_{\text{osc}}&= \left( \prod_{n_0+n_1>0 }
\int_{\mathbb{C}}d\phi_{n_0,n_1}d(\phi_{n_0,n_1})^*\right)e^{i S_{\text{eff}}[\phi_{\text{osc}},0]}\nonumber \\
&=\prod_{ n_0+n_1>0} \, {\frac{2i}{  (n_0+n_1) (n_1-n_0) - \frac{1}{{\mathcal M}} \big( n_0 + n_1 )^2 }}\;,
\label{InfProd}
\end{align}
which by letting $n:=n_0+n_1 $ and $m := n_1$ can be tidied up to
\begin{align}
Z_{\text{osc}}&=\prod_{\substack{m\\ n>0}}\, {\frac{i}{  n(m+Tn)}}\;,
\label{zetareg}
\end{align}
where
\begin{align}
\label{Tparameter}
    T :=-\frac12 (1+ {\mathcal M}^{-1})= - \frac{1}{1+ \tau}\;.
\end{align}

\subsection{Winding Modes}\label{app:wind}

Evaluating the effective action \eqref{Sen'sEffective} on the winding modes returns
\begin{align}
S_{\text{eff}}[\phi_{\text{w.m.}},J^\pm]  &=  \pi \left(   -R^2m_0^2+R^2m_1^2 -  \frac{R^2}{{\mathcal M}}(m_0+m_1)^2 
+2R(m_0+m_1)\mathcal{J}^{(0)} \right)\;,
\end{align}
where $\mathcal{J}^{(0)}$ is the normalised complex-structure zero mode defined in \eqref{zeromain}. Following on from the oscillator discussion we find
\begin{align}
Z_{\text{w.m.}} &= \sum_{\substack{m_0,m_1}} e^{i S_{\text{eff}}[\phi_{\text{w.m.}},J^\pm]}  \nonumber\\
&= \sum_{\substack{m_0,m_1}} e^{-i\pi R^2\left((m_0+m_1)(m_0-m_1)+(m_0+m_1)^2{\mathcal M}^{-1}\right) +2i\pi R (m_0+m_1)\mathcal{J}^{(0)}}\;.
\end{align} By introducing $n=m_0+m_1$ and $m=m_1$ we recast this into
\begin{align}\label{Zsumn}
Z_{\text{w.m.}} &= \sum_{\substack{m,n}}  e^{-i\pi R^2\left(n(n-2m)+n^2{\mathcal M}^{-1}\right) +2i\pi R n\mathcal{J}^{(0)}}\nonumber\\
&= \sum_{\substack{m,n}} e^{2\pi i R^2mn}e^{-i\pi R^2n^2\left(1+{\mathcal M}^{-1}\right) +2i\pi R n\mathcal{J}^{(0)}}\ .
\end{align}
The sum over $m$ is of the form
\begin{align}\label{deltasum}
\sum_{m\in\mathbb{Z}} e^{2 \pi i x m} = \sum_{q\in\mathbb{Z}} \delta(x - q)\ ,
\end{align}
which inserted into \eqref{Zsumn} gives
\begin{align}\label{delta}
Z_{\text{w.m.}} &= \sum_{n} \sum_{q}\delta(R^2n-q) e^{2i\pi R^2n^2T +2i\pi R n\mathcal{J}^{(0)}}\;,
\end{align}
where $T$ was defined in \eqref{Tparameter}. As discussed in Section \ref{sec:osc}, when $R^2$ is rational and equal to $r_1/r_2$ for $r_1$, $r_2$ coprime, we find contributions from $q=r_1p$, $n=r_2 p$:
\begin{align}\label{ZwmFinal}
    Z_{\text{w.m.}}&= \delta(0) \sum_{p} e^{2i\pi r_1r_2 p^2 T +2i\pi \sqrt{r_1r_2} p\mathcal{J}^{(0)}} ,
\end{align}
whereas when $R^2$ is irrational we find
\begin{align}
    Z_{\text{w.m.}}&= \delta(0)\ ,
\end{align}
as the sum  in (\ref{delta}) only has contributions from $n = q = 0$.

We want to understand the origin of the $\delta(0)$ divergence. 
In the case of an irrational $R^2$  it is clear that in terms of (\ref{Zsumn}) the $\delta(0)$ divergence arises from the infinite number of degenerate contributions to the partition function that  arise from chiral modes in the sum over $m$ at $n=0$.

In the case of rational $R^2$  we note that there is a  degeneracy where we shift a given winding mode $(m_0, m_1)$ by a chiral mode, $(s,- r_2 s), \ s \, \in \, \mathbb{Z}$. Despite  not being linear in the winding modes, one sees that upon shifting $\phi_{\text{w.m.}}\to \phi_{\text{w.m.}}+ \frac{R}{l}(s x^0-s x^1)$ the action transforms as
 \begin{align}
 S_{\text{eff}}[\phi_{\text{w.m.}},J^\pm]\to S_{\text{eff}}[\phi_{\text{w.m.}},J^\pm] - 2\pi r_1 p s\ ,
 \end{align}
 and hence $e^{i S_{\text{eff}}}$ is invariant. Thus summing over all winding modes induces a divergent contribution arising from an infinite number of states which only differ by a  chiral mode of $\phi$ but which all give the same contribution to the partition function. In essence this $\delta(0)$ therefore represents the contribution of the unphysical chiral modes that originate from the wrong-sign kinetic terms in the Lagrangian. We accordingly simply discard the infinite constant in Eq.~\eqref{Zwindingmain}.

\subsection{General Theta-Function Characteristics}\label{app:general}

To include general theta characteristics we take
\begin{align}
\phi_{\text{w.m.}} = \frac{R}{l}(m_0+\alpha_0)x^0 + 	\frac{R}{l}(m_1+\alpha_1)x^1\ ,
\end{align}
for some constants $\alpha_0,\alpha_1$. This will have the effect of shifting the sum over $n$ in (\ref{Zsumn}) to $n+\alpha_0+\alpha_1$  and thereby introducing the $\alpha$-characteristic in \eqref{thetas}.

In attempting to repeat the winding-mode calculation with the more general modings $m_0\to m_0+\alpha_0$, $m_1\to m_1+\alpha_1$, one quickly discovers that the sum (\ref{deltasum}) will now involve $x= R^2(n+\alpha_0+\alpha_1) $ and for generic $\alpha_0,\alpha_1$ this can never be integer (including zero). 
To counter this we add the following term to the action \eqref{eq:Sfull}:
\begin{align}\label{topterm}
S_{\cal A} :=  S - \frac{1}{2\pi}\int_{\mathbb T^2} {\cal A}	\wedge d\phi\ .
\end{align}
This term is a total derivative  and hence does not affect the equations of motion or any of the symmetries, including infinitesimal diffeomorphisms. As discussed around \eqref{Feyn}, we will bypass the Dirac path integral and work directly with the Feynman path-integral formulation of \eqref{eq:Sfull}, in terms of which one finds that the above term carries through and appears as is in the effective action \eqref{Sen'sEffective}. However, it does give the following contribution on the winding modes
\begin{align}
S_{{\cal A}\text{w.m.}} & =  S_{\text{w.m.}} - 2\pi R^2 \alpha_0(m_1+\alpha_1)	+2\pi R^2\alpha_1(m_0+\alpha_0)\nonumber\\
& =  S_{\text{w.m.}} - 2\pi R^2 (\alpha_0+\alpha_1 )m_1 + 2\pi R^2\alpha_1(m_0+m_1)\nonumber\\
& =  S_{\text{w.m.}} - 2\pi R^2 (\alpha_0+\alpha_1 )m + 2\pi R^2\alpha_1n
\ .
\end{align}
Performing the shift $n\to n +\alpha_0+\alpha_1$, $m\to m+\alpha_1$ in 
(\ref{Zsumn}) and including this extra term we find
\begin{align}
Z_{\text{w.m.}}
&= \sum_{n,m} e^{2\pi i R^2(m+\alpha_1)(n+\alpha_0+\alpha_1)}e^{2i\pi R^2(n+\alpha_0+\alpha_1)^2T +2i\pi R (n+\alpha_0+\alpha_1)\mathcal{J}^{(0)}} e^{-2\pi i R^2 (\alpha_0+\alpha_1 )m + 2\pi iR^2\alpha_1n}
\nonumber\\
& =  \sum_{n,m} e^{2\pi i R^2mn}e^{2i\pi R^2(n+\alpha_0+\alpha_1)^2T +2\pi i R (n+\alpha_0+\alpha_1)\mathcal{J}^{(0)}}  e^{2\pi i R^2 \alpha_1(\alpha_0+\alpha_1 ) + 4\pi iR^2\alpha_1n}\ .
\end{align}
The sum over $m$ reproduces the same $\delta$-function as in \eqref{delta}. Thus if $R^2=r_1/r_2$  we once again find $n=pr_2$ and hence
\begin{align}
Z_{\text{w.m.}}
&\sim e^{-\pi i\alpha\beta} \sum_p 	e^{2i\pi r_1r_2(p+\alpha)^2T +2i\pi  (p+\alpha)(\sqrt{r_1r_2}\mathcal{J}^{(0)}+\beta)}  \nonumber\\
& = e^{-\pi i\alpha\beta}\theta \! \begin{bmatrix} \vspace{-20pt} \\ \alpha \\ \vspace{-20pt} \\ \beta \\ \vspace{-20pt} \end{bmatrix} \! ( \sqrt{r_1r_2}\mathcal{J}^{(0)} |2 r_1r_2T )\ ,
\end{align}
where
\begin{align}
\alpha &= (\alpha_0+\alpha_1)/r_2\;,	\qquad \beta = 2r_1\alpha_1\ .
\end{align}

\subsection{Holomorphic Structure}\label{app:holo}
  
Consider the zero-mode contribution to the partition function \eqref{370main}. By observing that theta functions satisfy, for $m,n\in{\mathbb Z}$,
\begin{align}
    \theta \! \begin{bmatrix} \vspace{-20pt} \\ \alpha  \\ \vspace{-20pt} \\ \beta \\ \vspace{-20pt} \end{bmatrix} \! ( z + m +n\tau  | \tau ) = e^{-in^2\tau + 2\pi i(m\alpha-n\beta)-2\pi i n z} \theta \! \begin{bmatrix} \vspace{-20pt} \\ \alpha  \\ \vspace{-20pt} \\ \beta \\ \vspace{-20pt} \end{bmatrix} \! ( z    | \tau )\ ,
\end{align}
one sees that under a shift
\begin{align}\label{IntJ}
{\cal J}^{(0)} \to   {\cal J}^{(0)}  + \frac{m}{\sqrt{r_1r_2}} + 2n\sqrt{r_1r_2}T \ ,
\end{align}
the theta function in \eqref{370main} changes by:
\begin{align}
\theta \! \begin{bmatrix} \vspace{-20pt} \\ \alpha  \\ \vspace{-20pt} \\ \beta \\ \vspace{-20pt} \end{bmatrix} \! ( \sqrt{r_1r_2}{\cal J}^{(0)}    | 2r_1r_2T ) \to  e^{- 2 r_1 r_2 T n^2 + 2 \pi i (m \alpha - n \beta) - 2 \pi i \sqrt{r_1 r_2} \mathcal{J}^{(0)} n    }\theta \! \begin{bmatrix} \vspace{-20pt} \\ \alpha  \\ \vspace{-20pt} \\ \beta \\ \vspace{-20pt} \end{bmatrix} \! ( \sqrt{r_1r_2}{\cal J}^{(0)}    | 2r_1r_2T ) \ .    
\end{align}  
This is clearly a function of $T$ for any $m , n \neq 0$ ($\mathcal{J}^{(0)}$ itself has $T$ dependence) and since $T$ is complex this is not a pure phase. To see how the partition function transforms, we must also calculate the change to the anomalous prefactor $\mathcal{W}$. Shifting $\mathcal{J}^{(0)}$ by \eqref{IntJ} is equivalent to shifting the components of $J$ by
\begin{align}
    \begin{split}
    J^{+} &\to J^{+} + \frac{1}{2 l} \left( \frac{1}{\sqrt{r_1 r_2}}m - \sqrt{r_1 r_2} n  \right) \\
    J^{-} &\to J^{-} - \frac{1}{2 l} \sqrt{r_1 r_2} n \;,
    \end{split}
\end{align}
which can in turn be written as 
\begin{align}\label{lambdaxfm}
    J \to J + d \Lambda
    \;, \quad\textrm{with}\qquad \Lambda = \frac{1}{2l\sqrt{r_1r_2}}(m -2r_1r_2n)x^0- \frac{1}{2l}\frac{m}{\sqrt{r_1r_2}}x^1\ .
\end{align}
This can then be used to find the change in the anomalous term
\begin{align}
    \mathcal{W} \to e^{ 2\pi i r_1 r_2 T n^2 + 2 \pi i \sqrt{r_1 r_2} \mathcal{J}^{(0)}n + \frac{i}{2\pi} \int J \wedge d \Lambda  + \pi i m n }\, \mathcal{W} \;,
\end{align}
such that the overall change to $Z$ is simply
\begin{align} \label{phase}
    Z \to e^{\pi imn+ 2\pi i (m \alpha-n\beta))}  e^{\frac{i}{2 \pi} \int J \wedge d\Lambda} Z\;.
\end{align}
The change in the partition function is therefore a pure phase for all $m, n$. 

We observe that Eq.~\eqref{lambdaxfm} can be seen as part of the transformation (\ref{Lsym}). However, in this interpretation the winding modes of $\phi$ are similarly shifted by (\ref{Lsym}):
\begin{align}
m_0\to m_0  +     \frac{1}{2r_1}(m -2r_1r_2n)\;,\qquad m_1 \to m_1 -\frac{1}{2}\frac{m}{r_1}\;.
\end{align}
In order for this shift to make sense, {\it i.e.} for the winding modes to be mapped to winding modes, we see that we must restrict  $m=2r_1m'$ with $m'\in\mathbb Z$. Thus the transformation \eqref{Lsym} can only account for some of the shift symmetry of the partition function.

The action is not invariant under \eqref{Lsym}, but the change only depends on the sources and theta-characteristics.  Explicitly, we find that, when evaluated on  a winding mode $\phi_{\text{w.m.}}=R(m_\mu+\alpha_\mu)x^\mu/l$, the action shifts by\footnote{Note that there is an additional contribution to (\ref{SchangeSources}) arising from the ${\cal A}\wedge d\phi$ term.}
\begin{align}
S &\to S + \frac{1}{2\pi}\int d\Lambda\wedge (d\phi+{\cal A}-J)  \nonumber\\
& = S +\frac{\pi}{r_2}((m -2r_1r_2n)(m_1+2\alpha_1)+m(m_0+2\alpha_0))
-\frac{2l}{\sqrt{r_1r_2}}(r_1r_2 n J_+^{(0)} -(m-r_1r_2n)J_-^{(0)})\nonumber\\
& = S + \frac{2\pi}{r_2}(m(\alpha_0+\alpha_1)- 2r_1r_2n\alpha_1)-2l\sqrt{r_1r_2} n(-J_+^{(0)}+J_-^{(0)})+\frac{2l}{\sqrt{r_1r_2}} m J_-^{(0)}\nonumber\\
&\qquad +\frac{\pi}{r_2}m(m_1+m_0)- \pi  n m  \ .
\end{align}
To remove the dependence on the winding-mode numbers $m_0,m_1$ we see that in addition to $m=2r_1m'$, which makes the last term a multiple of $2\pi$, we also require $m'=r_2m''$ so that the second to last term is also a multiple of $2\pi$.  Thus for $m=2r_1r_2m''$ the shift in the action, modulo $2\pi$, is independent of the winding modes and hence we find that 
\begin{align}
Z\to e^{\frac{2\pi}{r_2}(m(\alpha_0+\alpha_1)- 2r_1r_2n\alpha_1)-2l\sqrt{r_1r_2} n(-J_+^{(0)}+J_-^{(0)})+\frac{2l}{\sqrt{r_1r_2}} m J_-^{(0)}} Z \;,
\end{align}
which agrees with \eqref{phase} for the identifications $\alpha = (\alpha_0 + \alpha_1)/r_2$ and $\beta = 2 r_1 \alpha$.

\section{Zeta-regularised product}\label{sec:appen}

In this appendix we will detail some aspects of the regularisation that we used in Section~\ref{sec:osc}. In what follows $T= -\frac{1}{2}(1 + \frac{1}{\mathcal M})=T_1+i T_2$, with $T_1,T_2\in\mathbb{R}$ and $T_2>0$ as long as $\tau_2>0$, see \eqref{MwickRotated}.

The infinite product
\begin{align}
P=\prod_{\substack{n > 0 \\ m\in{\mathbb Z}}}\frac{1 }{n(m + n T)}\ ,
\label{formalProduct}
\end{align}
can be regularised in the following standard fashion. First we will consider the auxiliary sum\footnote{From now on, $\sum_{m}$ will be a shorthand for $\sum_{m\in\mathbb{Z}}$.}
\begin{align}
G(s,T) := \sum_{n=1}^\infty\sum_{m} \frac{1}{n^s (m+ n T)^s}\;,
\label{auxiliarySum}
\end{align}
which is naturally defined for $\text{Re}(s)>1$. Then, by freely commuting the infinite sums with each other, and with the integrals that  appear,  we will analytically continue the latter to $\text{Re}(s)\ge0$. Finally, we will define \eqref{formalProduct} by
\begin{align}
P:=\exp\left[\frac{d}{ds}\Bigg|_{s=0}G(s,T) \right]\quad.
\label{ProductRegularized}
\end{align}
 
 It is easy to see that 
\begin{align}
 {G}(s,T_1+1,T_2)= {G}(s,T_1,T_2)\ ,
\end{align}
so we can employ its discrete Fourier transformation $F(s,p,T_2)$ defined by
\begin{align}
F(s,p,T_2):=\int_0^1d\chi e^{2\pi i \chi p} G(s,\chi,T_2)\ ,
\end{align}
to recast $ {G}(s, T_1, T_2)$ as
\begin{align}
 {G}(s, T_1, T_2)& = \sum_p e^{-2 \pi i p T_1} F(s,p,T_2)\cr
&=\sum_p e^{-2 \pi i p T_1}\int_{0}^{1} d \chi 
\sum_{n=1 }^{\infty}\sum_{m} \frac{1}{n^s (m+ n\chi+i n T_2)^s} e^{ 2 \pi i p \chi}\;.
\end{align}
Let $k \in \mathbb{Z}$ and $ r \in \{0,\dots, n-1\}$ so that we can write $m = kn+r$. Then the last line becomes
\begin{align}
 {G}(s, T_1, T_2) &= \sum_p \sum_{n = 1}^{\infty} \sum_{r = 0}^{n-1} \sum_k \int_{0}^{1} d \chi e^{2\pi i p ( \chi-T_1)}\frac{1}{n^s}\frac1{( r + n (\chi+k) +i nT_2)^s} \cr &= 
 \sum_p \sum_{n = 1}^{\infty} \sum_{r = 0}^{n-1} \sum_k \int_k^{k+1}d y e^{2\pi i p(y-T_1) }\frac{1}{n^s}\frac1{( r + n y + i n T_2)^s} \cr
 &=\sum_p \sum_{n = 1}^{\infty} \sum_{r = 0}^{n-1} \int_{-\infty}^{+\infty}d\chi e^{2\pi i p(\chi-\frac{r}{n}-T_1) }\frac1{\left[n^2( \chi  +i T_2)\right]^s} \cr
 &=\sum_k \sum_{n = 1}^{\infty}  \int_{-\infty}^{+\infty}d\chi e^{2\pi i kn(\chi-T_1) }n\frac1{\left[n^2( \chi  +i T_2)\right]^s} \ ,
 \label{afterR}
\end{align}
where we performed the change of coordinates $y:=\chi+k$ and $\chi:=y+\frac{r}{n}$ respectively in the second and third line, whereas in the last step we used
\begin{align}
\sum_{r = 0}^{n-1} e^{-2\pi i \frac{p}{n}r}=
\begin{cases}
n \quad  \text{if}\,\,\,p=kn\,\,\,k\in\mathbb{Z}\\
0 \quad \text{otherwise}
\end{cases}\,.
\end{align}
Since $T_2>0$ we can now implement the following integral representation of $z^{-s}$
\begin{align}
\frac{1}{z^s} = \frac{1}{i^s}\frac{1}{\Gamma(s)} \int_{0}^{\infty} d t \, t^{s-1} e^{i z t} \,\,\,\,\,\,\text{for }\,\,\, {\rm Im}(z)>0 \ ,
\end{align}
and, by switching to the $y:=n^2t$ variable, \eqref{afterR} becomes
\begin{align}
 {G}(s, T_1, T_2) &=\frac{1}{i^s}\frac{1}{\Gamma(s)} \sum_k \sum_{n = 1}^{\infty} n \int_{-\infty}^{+\infty}d\chi e^{2\pi i kn(\chi-T_1) }\int_{0}^{\infty} d t  t^{s-1} e^{in^2( \chi  +i T_2)t}\cr
&=\frac{1}{i^s}\frac{1}{\Gamma(s)} \sum_k \sum_{n = 1}^{\infty} \frac{1}{n^{2s-1}}\int_{0}^{\infty} dy  y^{s-1} e^{-2\pi i k n T_1-T_2y}\int_{-\infty}^{+\infty}d\chi e^{i\chi(2\pi kn+y) } \cr
&=\frac{2\pi}{i^s}\frac{1}{\Gamma(s)} \sum_{k=0} \sum_{n = 1}^{\infty} \frac{1}{n^{2s-1}}\int_{0}^{\infty} dy  y^{s-1} e^{-2\pi i k n T_1-T_2y}\delta(2\pi kn+y)  \cr
&=G_0(s,T_1,T_2)+
\frac{1}{i^s}\frac{1}{\Gamma(s)} \sum_{k=1}^{\infty}\sum_{n = 1}^{\infty} \left(\frac{2\pi k}{n}\right)^s \frac{e^{2\pi i k n T}}{k}  \; .
\label{CompactSeries}
\end{align}
Here we have split the $k=0$ contribution $G_0(s,T_1,T_2)$, which requires additional regularisation, from the $k>0$ terms which are instead convergent for $\text{Re}(s)\ge0$.  

To compute $G_0(s,T_1,T_2)$ we first observe that it is formally given by 
\begin{align}
  G_0(s,T_1,T_2) &=    \frac{2\pi}{i^s}\frac{1}{\Gamma(s)}   \sum_{n = 1}^{\infty} \frac{1}{n^{2s-1}}\int_{0}^{\infty} d y  y^{s-1} e^{-2\pi i k n T_1-T_2y}\delta(y)\nonumber\\
  & = \frac12 \frac{1}{i^s}\frac{1}{\Gamma(s)}  \sum_{n = 1}^{\infty}\left. \left(\frac{2\pi k}{n}\right)^s\frac{e^{2\pi i k n T}}{k}  \right|_{k=0}\ ,
\end{align}
where the extra factor of $1/2$ arises since
\begin{align}
\int_a^b d x\delta(x-c)f(x)=
\begin{cases}
f(c)\quad  \text{if}\,\,\,c\in (a,b)\\
\frac12 f(c)\quad  \text{if}\,\,\,c\in\{a,b\}\\
0 \quad \text{otherwise}
\end{cases}\,.
\end{align}
To regularise this  we
deform $k\to k+ \epsilon$   so that
\begin{align}
G_0(s,T_1,T_2) = \frac{1}{2}\frac{1}{i^s}\frac{1}{\Gamma(s)} \sum_{n = 1}^{\infty} \left(\frac{2\pi \epsilon}{n}\right)^s \frac{e^{2\pi i \epsilon n T}}{\epsilon}	
\end{align}
and therefore, for small $s$, 
\begin{align}
G_0(s,T_2,T_2)&=\frac{1}{2\Gamma(s)}\sum_{n = 1}^{\infty} \left(\frac{2\pi \epsilon}{in}\right)^s \frac{e^{2\pi i \epsilon n T}}{\epsilon} \nonumber\\
&= \frac{s}{2}\sum_{n = 1}^{\infty}\frac{e^{2\pi i \epsilon n T}}{\epsilon}+O(s^2)\nonumber\\
& =\frac{s}{2\epsilon}\zeta(0)+\pi i s T \zeta (-1)+O(s^2,\epsilon)\;,
\label{LastZeta}
\end{align}
where $\zeta(s)=\sum_{n=1}^{\infty}n^{-s}$ is the Riemann zeta function, for which $\zeta(0)=-\frac12$ and $\zeta(-1)=-\frac1{12}$. Since the divergent part of \eqref{LastZeta} does not depend on $T$, we regularise the $k=0$ contribution in \eqref{CompactSeries} by neglecting the $\frac1\epsilon$ divergence. In this case, when $s\to 0^+$, the expression \eqref{CompactSeries} looks like\footnote{We remind the reader that $\Gamma(s)=\frac1s-\gamma+O(s)$ where $\gamma$ is the Euler--Mascheroni constant.}
\begin{align}
{G}(s, T_1, T_2) &=s \pi i T \zeta(-1)+s \sum_{k=1}^{\infty} \sum_{n = 1}^{\infty} \left(\frac{2\pi k}{n}\right)^s \frac{e^{2\pi i k n T}}{k}  +O(s^2)\cr
&=s \pi i T \zeta(-1)-s \sum_{n = 1}^{\infty}\log(1-e^{2\pi i T n})  +O(s^2)\ ,
\label{ExpansionSsmall}
\end{align}
where we identified  $\sum_{k=1}^{\infty}\frac{q^k}{k}=-\log(1-q)$.
Recalling that $\zeta(-1)=-1/12$ we find
\begin{align}
P=\prod_{n=1}^\infty\prod_{m}\frac{1 }{n(m + n T)}= e^{-i\frac{\pi}{12}T} \prod_{n=1}^{\infty}\frac1{1-e^{2\pi i T n}} =\frac1{\eta(T) }\quad.
\end{align}
Finally, to make contact with the path-integral computations performed in Section~\ref{sec:osc} (see \eqref{zetaregmain}) we note that in our regularisation the product (let $a$ be a complex number)
\begin{align}
P_a=\prod_{\substack{n \neq 0 \\ m}}\frac{a }{n(m + n T)}\ ,
\end{align}
becomes 
\begin{align}
P_a=\exp\left[\frac{d}{ds}\Bigg|_{s=0}(a^s G(s,T)) \right]=a^{G(0,T)}\frac1{ \eta(T)  }\ ,
\end{align}
and by using \eqref{ExpansionSsmall} we find that $G(0,T)=0$.

 \end{appendix}

\bibliographystyle{utphys}
\bibliography{ref.bib}

\providecommand{\href}[2]{#2}\begingroup\raggedright\begin{thebibliography}{10}

\bibitem{Sen:2015nph}
A.~Sen, ``{Covariant Action for Type IIB Supergravity},''
  \href{http://dx.doi.org/10.1007/JHEP07(2016)017}{{\em JHEP} {\bfseries 07}
  (2016) 017}, \href{http://arxiv.org/abs/1511.08220}{{\ttfamily
  arXiv:1511.08220}}.

\bibitem{Sen:2019qit}
A.~Sen, ``{Self-dual forms: Action, Hamiltonian and Compactification},''
  \href{http://dx.doi.org/10.1088/1751-8121/ab5423}{{\em J. Phys.} {\bfseries
  A53} no.~8, (2020) 084002},
\href{http://arxiv.org/abs/1903.12196}{{\ttfamily arXiv:1903.12196 [hep-th]}}.

\bibitem{Siegel:1983es}
W.~Siegel, ``{Manifest Lorentz Invariance Sometimes Requires Nonlinearity},''
  \href{http://dx.doi.org/10.1016/0550-3213(84)90453-X}{{\em Nucl. Phys. B}
  {\bfseries 238} (1984) 307--316}.

\bibitem{Floreanini:1987as}
R.~Floreanini and R.~Jackiw, ``{Selfdual Fields as Charge Density Solitons},''
  \href{http://dx.doi.org/10.1103/PhysRevLett.59.1873}{{\em Phys. Rev. Lett.}
  {\bfseries 59} (1987) 1873}.

\bibitem{Imbimbo:1987yt}
C.~Imbimbo and A.~Schwimmer, ``{The Lagrangian Formulation of Chiral
  Scalars},'' \href{http://dx.doi.org/10.1016/0370-2693(87)91696-0}{{\em Phys.
  Lett. B} {\bfseries 193} (1987) 455--458}.

\bibitem{Bernstein:1988zd}
M.~Bernstein and J.~Sonnenschein, ``{A Comment on the Quantization of Chiral
  Bosons},'' \href{http://dx.doi.org/10.1103/PhysRevLett.60.1772}{{\em Phys.
  Rev. Lett.} {\bfseries 60} (1988) 1772}.

\bibitem{Henneaux:1988gg}
M.~Henneaux and C.~Teitelboim, ``{Dynamics of Chiral (Selfdual) $P$ Forms},''
\href{http://dx.doi.org/10.1016/0370-2693(88)90712-5}{{\em Phys. Lett.}
  {\bfseries B206} (1988) 650--654}.

\bibitem{McClain:1990sx}
B.~McClain, F.~Yu, and Y.~S. Wu, ``{Covariant quantization of chiral bosons and
  OSp(1,1|2) symmetry},''
\href{http://dx.doi.org/10.1016/0550-3213(90)90585-2}{{\em Nucl. Phys.}
  {\bfseries B343} (1990) 689--704}.

\bibitem{Pasti:1995tn}
P.~Pasti, D.~P. Sorokin, and M.~Tonin, ``{Duality symmetric actions with
  manifest space-time symmetries},''
  \href{http://dx.doi.org/10.1103/PhysRevD.52.R4277}{{\em Phys. Rev.}
  {\bfseries D52} (1995) R4277--R4281},
\href{http://arxiv.org/abs/hep-th/9506109}{{\ttfamily arXiv:hep-th/9506109
  [hep-th]}}.

\bibitem{Pasti:1995us}
P.~Pasti, D.~P. Sorokin, and M.~Tonin, ``{Space-time symmetries in duality
  symmetric models},'' in {\em {Gauge theories, applied supersymmetry, quantum
  gravity. Proceedings, Workshop, Leuven, Belgium, July 10-14, 1995}},
  pp.~167--176.
\newblock 1995.
\newblock
\href{http://arxiv.org/abs/hep-th/9509052}{{\ttfamily arXiv:hep-th/9509052
  [hep-th]}}.
\newblock

\bibitem{Perry:1996mk}
M.~Perry and J.~H. Schwarz, ``{Interacting chiral gauge fields in
  six-dimensions and Born-Infeld theory},''
  \href{http://dx.doi.org/10.1016/S0550-3213(97)00040-0}{{\em Nucl. Phys.}
  {\bfseries B489} (1997) 47--64},
\href{http://arxiv.org/abs/hep-th/9611065}{{\ttfamily arXiv:hep-th/9611065
  [hep-th]}}.

\bibitem{Pasti:1996vs}
P.~Pasti, D.~P. Sorokin, and M.~Tonin, ``{On Lorentz invariant actions for
  chiral p forms},'' \href{http://dx.doi.org/10.1103/PhysRevD.55.6292}{{\em
  Phys. Rev.} {\bfseries D55} (1997) 6292--6298},
\href{http://arxiv.org/abs/hep-th/9611100}{{\ttfamily arXiv:hep-th/9611100
  [hep-th]}}.

\bibitem{Pasti:1997gx}
P.~Pasti, D.~P. Sorokin, and M.~Tonin, ``{Covariant action for a D = 11
  five-brane with the chiral field},''
  \href{http://dx.doi.org/10.1016/S0370-2693(97)00188-3}{{\em Phys. Lett.}
  {\bfseries B398} (1997) 41--46},
\href{http://arxiv.org/abs/hep-th/9701037}{{\ttfamily arXiv:hep-th/9701037
  [hep-th]}}.

\bibitem{Aganagic:1997zq}
M.~Aganagic, J.~Park, C.~Popescu, and J.~H. Schwarz, ``{World volume action of
  the M theory five-brane},''
  \href{http://dx.doi.org/10.1016/S0550-3213(97)00227-7}{{\em Nucl. Phys.}
  {\bfseries B496} (1997) 191--214},
\href{http://arxiv.org/abs/hep-th/9701166}{{\ttfamily arXiv:hep-th/9701166
  [hep-th]}}.

\bibitem{Witten:1997sc}
E.~Witten, ``{Solutions of four-dimensional field theories via M theory},''
  \href{http://dx.doi.org/10.1016/S0550-3213(97)00416-1}{{\em Nucl. Phys.}
  {\bfseries B500} (1997) 3--42},
  \href{http://arxiv.org/abs/hep-th/9703166}{{\ttfamily arXiv:hep-th/9703166}}.
  [,452(1997)].

\bibitem{Witten:1999vg}
E.~Witten, ``{Duality relations among topological effects in string theory},''
  \href{http://dx.doi.org/10.1088/1126-6708/2000/05/031}{{\em JHEP} {\bfseries
  05} (2000) 031},
\href{http://arxiv.org/abs/hep-th/9912086}{{\ttfamily arXiv:hep-th/9912086
  [hep-th]}}.

\bibitem{Belov:2006jd}
D.~Belov and G.~W. Moore, ``{Holographic Action for the Self-Dual Field},''
\href{http://arxiv.org/abs/hep-th/0605038}{{\ttfamily arXiv:hep-th/0605038
  [hep-th]}}.

\bibitem{Mkrtchyan:2019opf}
K.~Mkrtchyan, ``{On Covariant Actions for Chiral $p-$Forms},''
  \href{http://dx.doi.org/10.1007/JHEP12(2019)076}{{\em JHEP} {\bfseries 12}
  (2019) 076},
\href{http://arxiv.org/abs/1908.01789}{{\ttfamily arXiv:1908.01789 [hep-th]}}.

\bibitem{Lambert:2019diy}
N.~Lambert, ``{(2,0) Lagrangian Structures},''
  \href{http://dx.doi.org/10.1016/j.physletb.2019.134948}{{\em Phys. Lett.}
  {\bfseries B798} (2019) 134948},
  \href{http://arxiv.org/abs/1908.10752}{{\ttfamily arXiv:1908.10752}}.

\bibitem{Andriolo:2020ykk}
E.~Andriolo, N.~Lambert, and C.~Papageorgakis, ``{Geometrical Aspects of An
  Abelian (2,0) Action},''
  \href{http://dx.doi.org/10.1007/JHEP04(2020)200}{{\em JHEP} {\bfseries 04}
  (2020) 200}, \href{http://arxiv.org/abs/2003.10567}{{\ttfamily
  arXiv:2003.10567 [hep-th]}}.

\bibitem{Alvarez-Gaume:1986nqf}
L.~Alvarez-Gaume, G.~W. Moore, P.~C. Nelson, C.~Vafa, and J.~b. Bost,
  ``{Bosonization in Arbitrary Genus},''
  \href{http://dx.doi.org/10.1016/0370-2693(86)90466-1}{{\em Phys. Lett. B}
  {\bfseries 178} (1986) 41--47}.

\bibitem{Alvarez-Gaume:1987wwg}
L.~Alvarez-Gaume, J.~B. Bost, G.~W. Moore, P.~C. Nelson, and C.~Vafa,
  ``{Bosonization on Higher Genus Riemann Surfaces},''
  \href{http://dx.doi.org/10.1007/BF01218489}{{\em Commun. Math. Phys.}
  {\bfseries 112} (1987) 503}.

\bibitem{Henningson:1999dm}
M.~Henningson, B.~E.~W. Nilsson, and P.~Salomonson, ``{Holomorphic
  factorization of correlation functions in (4k+2)-dimensional (2k) form gauge
  theory},'' \href{http://dx.doi.org/10.1088/1126-6708/1999/09/008}{{\em JHEP}
  {\bfseries 09} (1999) 008},
  \href{http://arxiv.org/abs/hep-th/9908107}{{\ttfamily arXiv:hep-th/9908107}}.

\bibitem{Freed:2006ya}
D.~S. Freed, G.~W. Moore, and G.~Segal, ``{The Uncertainty of Fluxes},''
  \href{http://dx.doi.org/10.1007/s00220-006-0181-3}{{\em Commun. Math. Phys.}
  {\bfseries 271} (2007) 247--274},
  \href{http://arxiv.org/abs/hep-th/0605198}{{\ttfamily arXiv:hep-th/0605198}}.

\bibitem{Freed:2006yc}
D.~S. Freed, G.~W. Moore, and G.~Segal, ``{Heisenberg Groups and Noncommutative
  Fluxes},'' \href{http://dx.doi.org/10.1016/j.aop.2006.07.014}{{\em Annals
  Phys.} {\bfseries 322} (2007) 236--285},
  \href{http://arxiv.org/abs/hep-th/0605200}{{\ttfamily arXiv:hep-th/0605200}}.

\bibitem{Witten:1988hf}
E.~Witten, ``{Quantum Field Theory and the Jones Polynomial},''
  \href{http://dx.doi.org/10.1007/BF01217730}{{\em Commun. Math. Phys.}
  {\bfseries 121} (1989) 351--399}.

\bibitem{Witten:1996hc}
E.~Witten, ``{Five-brane effective action in M theory},''
  \href{http://dx.doi.org/10.1016/S0393-0440(97)80160-X}{{\em J. Geom. Phys.}
  {\bfseries 22} (1997) 103--133},
\href{http://arxiv.org/abs/hep-th/9610234}{{\ttfamily arXiv:hep-th/9610234
  [hep-th]}}.

\bibitem{Devecchi:1996cp}
F.~P. Devecchi and M.~Henneaux, ``{Covariant path integral for chiral p
  forms},'' \href{http://dx.doi.org/10.1103/PhysRevD.54.1606}{{\em Phys. Rev.}
  {\bfseries D54} (1996) 1606--1613},
\href{http://arxiv.org/abs/hep-th/9603031}{{\ttfamily arXiv:hep-th/9603031
  [hep-th]}}.

\bibitem{Chen:2013gca}
W.-M. Chen, P.-M. Ho, H.-c. Kao, F.~S. Khoo, and Y.~Matsuo, ``{Partition
  function of a chiral boson on a 2-torus from the
  Floreanini\textendash{}Jackiw Lagrangian},''
  \href{http://dx.doi.org/10.1093/ptep/ptu021}{{\em PTEP} {\bfseries 2014}
  no.~3, (2014) 033B02}, \href{http://arxiv.org/abs/1307.2172}{{\ttfamily
  arXiv:1307.2172 [hep-th]}}.

\bibitem{Visser:2017atf}
M.~Visser, ``{How to Wick rotate generic curved spacetime},''
  \href{http://arxiv.org/abs/1702.05572}{{\ttfamily arXiv:1702.05572 [gr-qc]}}.

\bibitem{Kontsevich:2021dmb}
M.~Kontsevich and G.~Segal, ``{Wick Rotation and the Positivity of Energy in
  Quantum Field Theory},'' \href{http://dx.doi.org/10.1093/qmath/haab027}{{\em
  Quart. J. Math. Oxford Ser.} {\bfseries 72} no.~1-2, (2021) 673--699},
  \href{http://arxiv.org/abs/2105.10161}{{\ttfamily arXiv:2105.10161
  [hep-th]}}.

\bibitem{Witten:2021nzp}
E.~Witten, ``{A Note On Complex Spacetime Metrics},''
  \href{http://arxiv.org/abs/2111.06514}{{\ttfamily arXiv:2111.06514
  [hep-th]}}.

\bibitem{Visser:2021ucg}
M.~Visser, ``{Feynman's i-epsilon prescription, almost real spacetimes, and
  acceptable complex spacetimes},''
  \href{http://arxiv.org/abs/2111.14016}{{\ttfamily arXiv:2111.14016 [gr-qc]}}.

\bibitem{Moore:1988qv}
G.~W. Moore and N.~Seiberg, ``{Classical and Quantum Conformal Field Theory},''
  \href{http://dx.doi.org/10.1007/BF01238857}{{\em Commun. Math. Phys.}
  {\bfseries 123} (1989) 177}.

\bibitem{Moore:1989yh}
G.~W. Moore and N.~Seiberg, ``{Taming the Conformal Zoo},''
  \href{http://dx.doi.org/10.1016/0370-2693(89)90897-6}{{\em Phys. Lett. B}
  {\bfseries 220} (1989) 422--430}.

\bibitem{Elitzur:1989nr}
S.~Elitzur, G.~W. Moore, A.~Schwimmer, and N.~Seiberg, ``{Remarks on the
  Canonical Quantization of the Chern-Simons-Witten Theory},''
  \href{http://dx.doi.org/10.1016/0550-3213(89)90436-7}{{\em Nucl. Phys. B}
  {\bfseries 326} (1989) 108--134}.

\bibitem{Vanichchapongjaroen:2020wza}
P.~Vanichchapongjaroen, ``{Covariant M5-brane action with self-dual 3-form},''
  \href{http://dx.doi.org/10.1007/JHEP05(2021)039}{{\em JHEP} {\bfseries 05}
  (2021) 039}, \href{http://arxiv.org/abs/2011.14384}{{\ttfamily
  arXiv:2011.14384 [hep-th]}}.

\bibitem{Senjanovic:1976br}
P.~Senjanovic, ``{Path Integral Quantization of Field Theories with Second
  Class Constraints},''
  \href{http://dx.doi.org/10.1016/0003-4916(76)90062-2}{{\em Annals Phys.}
  {\bfseries 100} (1976) 227--261}. [Erratum: Annals Phys. 209, 248 (1991)].

\bibitem{Henneaux:1992ig}
M.~Henneaux and C.~Teitelboim, {\em {Quantization of gauge systems}}.
\newblock Princeton University Press, 1992.

\bibitem{DiFrancesco:1997nk}
P.~Di~Francesco, P.~Mathieu, and D.~Senechal,
  \href{http://dx.doi.org/10.1007/978-1-4612-2256-9}{{\em {Conformal Field
  Theory}}}.
\newblock Graduate Texts in Contemporary Physics. Springer-Verlag, New York,
  1997.

\end{thebibliography}\endgroup

\end{document}